\documentclass[12pt,prd,showpacs,tightenlines,nofootinbib]{revtex4}
\usepackage{bm}
\usepackage{graphics}
\usepackage{rotating}
\usepackage{epsfig}
\begin{document}
\begin{flushright}{HU-EP-10-34}\end{flushright}
\title{Semileptonic and nonleptonic decays of $B_c$ mesons to 
orbitally excited heavy mesons in the relativistic quark model}
\author{D. Ebert$^{1}$, R. N. Faustov$^{1,2}$  and V. O. Galkin$^{1,2}$}
\affiliation{
$^1$ Institut f\"ur Physik, Humboldt--Universit\"at zu Berlin,
Newtonstr. 15, D-12489  Berlin, Germany\\
$^2$ Dorodnicyn Computing Centre, Russian Academy of Sciences,
  Vavilov Str. 40, 119991 Moscow, Russia}

\begin{abstract}
The form factors of weak decays of the $B_c$ meson
to orbitally excited charmonium, $D$, $B_s$ and $B$ mesons are
calculated in the framework of the QCD-motivated relativistic quark
model based on the quasipotential approach.  Relativistic effects are
systematically taken into account. The form factor dependence
on the  momentum transfer is reliably determined in the whole
kinematical range. The form factors are expressed trough the overlap
integrals of the meson wave functions which are known from the
previous mass spectra calculations within the same model. On this basis
semileptonic and nonleptonic $B_c$ decay rates to orbitally excited
heavy mesons are calculated. Predictions for the $B_c$ decays to the
orbitally and radially excited $2P$
and $3S$ charmonium states are given which could be used for clarifying the
nature of the recently observed charmonium-like states above the open charm
production threshold.   

\end{abstract}


\maketitle

\section{Introduction}
\label{sec:int}

The investigation of weak decays of mesons composed of a heavy quark and
antiquark gives a very important insight in the heavy quark dynamics.
The decay properties of the $B_c$ meson are of special interest, since it
is the only heavy meson consisting of two heavy quarks with different
flavor. This difference of quark flavors forbids annihilation into
gluons. As a result, the excited $B_c$ meson states lying below the
$BD$ meson threshold undergo pionic or radiative transitions to
the pseudoscalar ground state which is considerably more
stable than corresponding charmonium or bottomonium states and  decays
only weakly.  The CDF Collaboration reported the discovery of the
$B_c$ ground state in $p\bar p$ collisions already more than ten years
ago \cite{cdf}. However, up till recently its mass was known with a very large
error. Now it is measured with a good precision in the decay
channel $B_c\to J/\psi\pi$. The measured value
$M_{B_c}^{\rm\mbox{\scriptsize exp}}=6275.2\pm 
2.9\pm2.5$~MeV \cite{cdfm} is in a very good agreement with the
prediction of the relativistic quark model $M_{B_c}^{\rm\mbox{\scriptsize
theor}}=6270$~MeV \cite{mass}. More experimental data on masses and decays
of the $B_c$ mesons are expected to come in near future
from the Tevatron at Fermilab and the Large Hadron Collider (LHC) at CERN.

The characteristic feature of the $B_c$ meson  is that both
quarks forming it are heavy and thus their weak decays give comparable
contributions to the total decay rate. Therefore it is necessary to
consider both the $b$ quark transitions $b\to c,u$ with the $\bar c$ quark
being a spectator and $\bar c$ quark transitions $\bar c\to \bar s,\bar d$
with the $b$ quark being a
spectator. The former transitions lead to weak decays to
charmonium and $D$ mesons while the latter lead to decays to $B_s$ and
$B$ mesons. 
The estimates \cite{qwg} of the $B_c$ decay rates indicate that the $c$ quark transitions
give the dominant contribution ($\sim 70\%$) while the $b$ quark
transitions and weak annihilation contribute about 20\% and 10\%,
respectively. However, from the experimental point of view the $B_c$
decays to charmonium are easier to identify. Indeed, CDF and D0
observed the $B_c$ meson and measured its mass analyzing its
semileptonic and nonleptonic 
decays  $B_c\to J/\psi l\nu$ and $B_c\to J/\psi\pi$ \cite{cdf,cdfm,d0}. 

In this paper we  extend our previous investigation of $B_c$ properties
\cite{mass,bcjpsi,bcbs} to study exclusive weak semileptonic and nonleptonic
decay channels to orbitally excited heavy mesons. For the
calculations we use the same effective methods \cite{bcjpsi,bcbs} developed in the framework
of the relativistic quark model based on the quasipotential
approach for the $B_c$ decays to ground and radially excited states of
charmonium, $D$, $B_s$ and $B$ mesons. Here weak decays to 
orbital excitations of these mesons, governed both  by the $b$ and $c$ quark decays,
are considered.  The weak decay matrix elements are parametrized by
invariant form factors which are then expressed through the overlap integrals of
the meson wave functions. The systematic account for the relativistic
effects, including wave function transformations from the rest to the
moving frame and contributions from the intermediate negative-energy
states, allows to reliably determine the momentum transfer dependence
of the decay form factors in the whole accessible kinimatical range. The
other important advantage of our approach is that for numerical calculations we use
the relativistic wave functions, obtained in the
meson mass spectra calculations, and not some ad hoc parametrizations
which were widely used in some previous investigations. The calculated
form factors are then substituted in the expressions for the differential
decay rates. 

The important distinction between weak $B_c$ decays,
associated with the $b$ and $c$ quark decays, consists in the
significant difference of the accessible kinimatical ranges. In the $B_c$
decays to  the charmonium and $D$ mesons the kinimatical range is considerably broader
(by about an order of magnitude) than for decays to $B_s$ and $B$
mesons. As a result, many weak decays which are kinematically allowed in
the former case are forbidden in the latter one. The kinematical
suppression of semileptonic $B_c\to B_s(B)l\nu$ decays should be more
pronounced for the decays to excited states than for the ground
ones. The nonleptonic $B_c$ decays to an orbitally excited heavy meson
and an energetic light meson can then be considered on the basis of
the factorization approximation. The obtained predictions for the
decay rates are compared with previous calculations which are based on
different relativistic quark models \cite{iks,hnv,ccwz},      
three-point QCD sum rules \cite{asb} and light-cone QCD sum
rules \cite{wl}. 

We also consider here weak semileptonic and nonleptonic $B_c$ decays to the
highly excited $2P$ and $3S$ charmonium states. These states are of
special interest since in last years a number of new
charmonium-like states above the open charm production threshold have
been observed  \cite{pakhlova}. They include several unexpectedly narrow states,
$X(3872)$, $X(3940)$, $Y(3940)$, $Z(3930)$, $Y(4260)$, $Z(4430)$, which
interpretation is controversial. Some of them could
be candidates for excited charmonia. Therefore experimental
observation of such states in $B_c$ decays could help to clarify
their real nature.   

\section{Relativistic quark model}  
\label{rqm}

In the quasipotential approach a meson is described as a bound
quark-antiquark state with a wave function satisfying the
quasipotential equation of the Schr\"odinger type 
\begin{equation}
\label{quas}
{\left(\frac{b^2(M)}{2\mu_{R}}-\frac{{\bf
p}^2}{2\mu_{R}}\right)\Psi_{M}({\bf p})} =\int\frac{d^3 q}{(2\pi)^3}
 V({\bf p,q};M)\Psi_{M}({\bf q}),
\end{equation}
where the relativistic reduced mass is
\begin{equation}
\mu_{R}=\frac{E_1E_2}{E_1+E_2}=\frac{M^4-(m^2_1-m^2_2)^2}{4M^3},
\end{equation}
and $E_1$, $E_2$ are the center of mass energies on mass shell given by
\begin{equation}
\label{ee}
E_1=\frac{M^2-m_2^2+m_1^2}{2M}, \quad E_2=\frac{M^2-m_1^2+m_2^2}{2M}.
\end{equation}
Here $M=E_1+E_2$ is the meson mass, $m_{1,2}$ are the quark masses,
and ${\bf p}$ is their relative momentum.  
In the center of mass system the relative momentum squared on mass shell 
reads
\begin{equation}
{b^2(M) }
=\frac{[M^2-(m_1+m_2)^2][M^2-(m_1-m_2)^2]}{4M^2}.
\end{equation}

The kernel 
$V({\bf p,q};M)$ in Eq.~(\ref{quas}) is the quasipotential operator of
the quark-antiquark interaction. It is constructed with the help of the
off-mass-shell scattering amplitude, projected onto the positive
energy states. 
Constructing the quasipotential of the quark-antiquark interaction, 
we have assumed that the effective
interaction is the sum of the usual one-gluon exchange term with the mixture
of long-range vector and scalar linear confining potentials, where
the vector confining potential
contains the Pauli interaction. The quasipotential is then defined by
\cite{mass}
  \begin{equation}
\label{qpot}
V({\bf p,q};M)=\bar{u}_1(p)\bar{u}_2(-p){\mathcal V}({\bf p}, {\bf
q};M)u_1(q)u_2(-q),
\end{equation}
with
$${\mathcal V}({\bf p},{\bf q};M)=\frac{4}{3}\alpha_sD_{ \mu\nu}({\bf
k})\gamma_1^{\mu}\gamma_2^{\nu}
+V^V_{\rm conf}({\bf k})\Gamma_1^{\mu}
\Gamma_{2;\mu}+V^S_{\rm conf}({\bf k}),$$
where $\alpha_s$ is the QCD coupling constant, $D_{\mu\nu}$ is the
gluon propagator in the Coulomb gauge
\begin{equation}
D^{00}({\bf k})=-\frac{4\pi}{{\bf k}^2}, \quad D^{ij}({\bf k})=
-\frac{4\pi}{k^2}\left(\delta^{ij}-\frac{k^ik^j}{{\bf k}^2}\right),
\quad D^{0i}=D^{i0}=0,
\end{equation}
and ${\bf k=p-q}$. Here $\gamma_{\mu}$ and $u(p)$ are 
the Dirac matrices and spinors
\begin{equation}
\label{spinor}
u^\lambda({p})=\sqrt{\frac{\epsilon(p)+m}{2\epsilon(p)}}
\left(
\begin{array}{c}1\cr {\displaystyle\frac{\bm{\sigma}
      {\bf  p}}{\epsilon(p)+m}}
\end{array}\right)\chi^\lambda,
\end{equation}
where  $\bm{\sigma}$   and $\chi^\lambda$
are Pauli matrices and spinors and $\epsilon(p)=\sqrt{{\bf p}^2+m^2}$.
The effective long-range vector vertex is
given by
\begin{equation}
\label{kappa}
\Gamma_{\mu}({\bf k})=\gamma_{\mu}+
\frac{i\kappa}{2m}\sigma_{\mu\nu}k^{\nu},
\end{equation}
where $\kappa$ is the Pauli interaction constant characterizing the
long-range anomalous chromomagnetic moment of quarks. Vector and
scalar confining potentials in the nonrelativistic limit reduce to
\begin{eqnarray}
\label{vlin}
V_V(r)&=&(1-\varepsilon)(Ar+B),\nonumber\\ 
V_S(r)& =&\varepsilon (Ar+B),
\end{eqnarray}
reproducing 
\begin{equation}
\label{nr}
V_{\rm conf}(r)=V_S(r)+V_V(r)=Ar+B,
\end{equation}
where $\varepsilon$ is the mixing coefficient. 

The expression for the quasipotential of the heavy quarkonia,
expanded in $v^2/c^2$  can be found in Ref.~\cite{mass}. The
quasipotential for the heavy quark interaction with a light antiquark
without employing the nonrelativistic ($v/c$)  expansion 
is given in Ref.~\cite{hlm}.  All the parameters of
our model like quark masses, parameters of the linear confining potential
$A$ and $B$, mixing coefficient $\varepsilon$ and anomalous
chromomagnetic quark moment $\kappa$ are fixed from the analysis of
heavy quarkonium masses and radiative
decays  \cite{mass}. The quark masses
$m_b=4.88$ GeV, $m_c=1.55$ GeV, $m_s=0.5$ GeV, $m_{u,d}=0.33$ GeV and
the parameters of the linear potential $A=0.18$ GeV$^2$ and $B=-0.30$ GeV
have values inherent for quark models.  The value of the mixing
coefficient of vector and scalar confining potentials $\varepsilon=-1$
has been determined from the consideration of the heavy quark expansion
for the semileptonic $B\to D$ decays
\cite{fg} and charmonium radiative decays \cite{mass}.
Finally, the universal Pauli interaction constant $\kappa=-1$ has been
fixed from the analysis of the fine splitting of heavy quarkonia ${
}^3P_J$- states \cite{mass} and  the heavy quark expansion for semileptonic
decays of heavy mesons \cite{fg} and baryons \cite{sbar}. Note that the 
long-range  magnetic contribution to the potential in our model
is proportional to $(1+\kappa)$ and thus vanishes for the 
chosen value of $\kappa=-1$ in accordance with the flux tube model.

\section{Matrix elements of the electroweak current for
  ${\lowercase{b\to c,u}}$ and ${\lowercase{c\to s,d}}$ transitions} \label{mml}

In order to calculate the exclusive semileptonic decay rate of the
$B_c$ meson, it is necessary to determine the corresponding matrix
element of the  weak current between meson states. First we consider
the weak $B_c$ decays governed by the $b$ quark decays. 
In the quasipotential approach,  the matrix element of the weak current
$J^W_\mu=\bar q\gamma_\mu(1-\gamma_5)b$, associated with the $b\to q$ ($q=c$
or $u$) transition, between a $B_c$ meson with mass $M_{B_c}$ and
momentum $p_{B_c}$ and a final $P$-wave meson $F$ ($F=\chi_{cJ},h_c$ or
$D_J^{(*)}$) with mass $M_F$ and momentum $p_F$ takes the form \cite{f}
\begin{equation}\label{mxet} 
\langle F(p_F) \vert J^W_\mu \vert B_c(p_{B_c})\rangle
=\int \frac{d^3p\, d^3q}{(2\pi )^6} \bar \Psi_{F\,{\bf p}_F}({\bf
p})\Gamma _\mu ({\bf p},{\bf q})\Psi_{B_c\,{\bf p}_{B_c}}({\bf q}),
\end{equation}
where $\Gamma _\mu ({\bf p},{\bf
q})$ is the two-particle vertex function and  
$\Psi_{M\,{\bf p}_M}$ are the
meson ($M=B_c,F)$ wave functions projected onto the positive energy states of
quarks and boosted to the moving reference frame with momentum ${\bf p}_M$.
\begin{figure}
  \centering
  \includegraphics{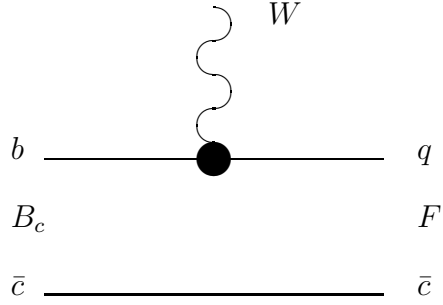}
\caption{Lowest order vertex function $\Gamma^{(1)}$
contributing to the current matrix element (\ref{mxet}). \label{d1}}
\end{figure}

\begin{figure}
  \centering
  \includegraphics{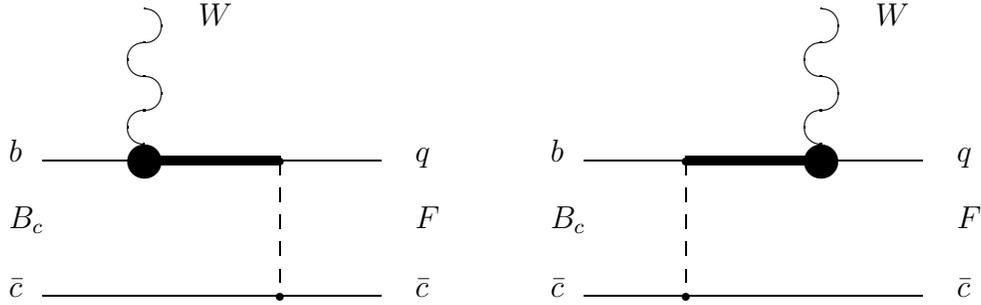}
\caption{ Vertex function $\Gamma^{(2)}$
taking the quark interaction into account. Dashed lines correspond  
to the effective potential ${\cal V}$ in 
(\ref{qpot}). Bold lines denote the negative-energy part of the quark
propagator. \label{d2}}
\end{figure}

 The contributions to $\Gamma$ come from Figs.~\ref{d1} and \ref{d2}. 
The contribution $\Gamma^{(2)}$ is the consequence
of the projection onto the positive-energy states. Note that the form of the
relativistic corrections emerging from the vertex function
$\Gamma^{(2)}$ explicitly depend on the Lorentz structure of the
quark-antiquark interaction. In the leading order of the $v^2/c^2$
expansion for 
$B_c$ and $\chi_J$ and in the heavy quark limit $m_{c}\to \infty$ for $D_J$
only $\Gamma^{(1)}$ contributes, while $\Gamma^{(2)}$  
contributes at the subleading order. 
The vertex functions look like
\begin{equation} \label{gamma1}
\Gamma_\mu^{(1)}({\bf
p},{\bf q})=\bar u_{q}(p_q)\gamma_\mu(1-\gamma^5)u_b(q_b)
(2\pi)^3\delta({\bf p}_c-{\bf
q}_c),\end{equation}
and
\begin{eqnarray}\label{gamma2} 
\Gamma_\mu^{(2)}({\bf
p},{\bf q})&=&\bar u_{q}(p_q)\bar u_c(p_c) \Bigl\{\gamma_{1\mu}(1-\gamma_1^5)
\frac{\Lambda_b^{(-)}(
k)}{\epsilon_b(k)+\epsilon_b(p_q)}\gamma_1^0
{\cal V}({\bf p}_c-{\bf
q}_c)\nonumber \\ 
& &+{\cal V}({\bf p}_c-{\bf
q}_c)\frac{\Lambda_{q}^{(-)}(k')}{ \epsilon_{q}(k')+
\epsilon_{q}(q_b)}\gamma_1^0 \gamma_{1\mu}(1-\gamma_1^5)\Bigr\}u_b(q_b)
u_c(q_c),\end{eqnarray}
where the superscripts ``(1)" and ``(2)" correspond to Figs.~\ref{d1} and
\ref{d2},  ${\bf k}={\bf p}_q-{\bf\Delta};\
{\bf k}'={\bf q}_b+{\bf\Delta};\ {\bf\Delta}={\bf
p}_F-{\bf p}_{B_c}$;
$$\Lambda^{(-)}(p)=\frac{\epsilon(p)-\bigl( m\gamma
^0+\gamma^0({\bm{ \gamma}{\bf p}})\bigr)}{ 2\epsilon (p)}.$$
Here \cite{f} 
\begin{eqnarray*} 
p_{q,c}&=&\epsilon_{q,c}(p)\frac{p_F}{M_F}
\pm\sum_{i=1}^3 n^{(i)}(p_F)p^i,\\
q_{b,c}&=&\epsilon_{b,c}(q)\frac{p_{B_c}}{M_{B_c}} \pm \sum_{i=1}^3 n^{(i)}
(p_{B_c})q^i,\end{eqnarray*}
and $n^{(i)}$ are three four-vectors given by
$$ n^{(i)\mu}(p)=\left\{ \frac{p^i}{M},\ \delta_{ij}+
\frac{p^ip^j}{M(E+M)}\right\}, \quad E=\sqrt{{\bf p}^2+M^2}.$$

The wave function of a final $P$-wave $F$ meson at rest is given by
\begin{equation}\label{psi}
\Psi_{F}({\bf p})\equiv
\Psi^{J{\cal M}}_{F(^{2S+1}P_J)}({\bf p})={\cal Y}^{J{\cal M}}_S\,\psi_{F(^{2S+1}P_J)}({\bf p}),
\end{equation}
where $J$ and ${\cal M}$ are the total meson angular momentum and its projection,
while $S=0,1$ is the total spin.   
$\psi_{F(^{2S+1}P_J)}({\bf p})$ is the radial part of the wave function,
which has been determined by the numerical solution of Eq.~(\ref{quas})
in \cite{mass,hlm}.
The spin-angular momentum part ${\cal Y}^{J{\cal M}}_S$ has the following form
\begin{equation}\label{angl}
{\cal Y}^{J{\cal M}}_S=\sum_{\sigma_1\sigma_2}\langle 1\, {\cal M}-\sigma_1-\sigma_2,\  
S\, \sigma_1+\sigma_2 |J\, {\cal M}\rangle\langle \frac12\, \sigma_1,\ 
\frac12\, \sigma_2 |S\, \sigma_1+\sigma_2\rangle Y_{1}^{{\cal M}-\sigma_1-\sigma_2}
\chi_1(\sigma_1)\chi_2(\sigma_2).
\end{equation}
Here $\langle j_1\, m_1,\  j_2\, m_2|J\, {\cal M}\rangle$ are the Clebsch-Gordan 
coefficients, $Y_l^m$ are spherical harmonics, and $\chi(\sigma)$ (where 
$\sigma=\pm 1/2$) are spin wave functions,
$$ \chi\left(1/2\right)={1\choose 0}, \qquad 
\chi\left(-1/2\right)={0\choose 1}. $$

The heavy-light meson states (such as $D_1$, $D_1'$ etc.)   with $J=L=1$
are  mixtures of spin-triplet $F(^3P_1)$  and spin-singlet $F(^1P_1)$
states:
\begin{eqnarray}
  \label{eq:mix}
  \Psi_{F_1}&=&\Psi_{F(^1P_1)}\cos\varphi+\Psi_{F(^3P_1)}\sin\varphi, \cr
 \Psi_{F_1'}&=&-\Psi_{F(^1P_1)}\sin\varphi+\Psi_{F(^3P_1)}\cos\varphi, 
\end{eqnarray}
where $\varphi$ is the mixing angle and the primed state has the heavier mass.
  Such mixing occurs due to the nondiagonal spin-orbit and
tensor terms in the $Q\bar q$ quasipotential. The physical states are obtained
by diagonalizing the corresponding mixing terms. The values of the mixing angle
$\varphi$ were determined in the heavy-light meson mass spectra
calculations \cite{hlm} and are given in Table~\ref{tab:mix}. 
\begin{table}
\caption{Mixing angles $\varphi$ for heavy-light mesons (in $^{\circ}$).} 
   \label{tab:mix}
\begin{ruledtabular}
\begin{tabular}{ccccc}
State& $D$ & $D_s$ & $B$ & $B_s$ \\
\hline
$1P$ & 35.5 & 34.5 & 35.0 & 36.0\\
$2P$ & 37.5 & 37.6 & 37.3 & 34.0\\
\end{tabular}
\end{ruledtabular}
\end{table}

It is important to note that the wave functions entering the weak current
matrix element (\ref{mxet}) are not in the rest frame in general. For example, 
in the $B_c$ meson rest frame (${\bf p}_{B_c}=0$), the final  meson
is moving with the recoil momentum ${\bf \Delta}$. The wave function
of the moving  meson $\Psi_{F\,{\bf\Delta}}$ is connected 
with the  wave function in the rest frame 
$\Psi_{F\,{\bf 0}}\equiv \Psi_F$ by the transformation \cite{f}
\begin{equation}
\label{wig}
\Psi_{F\,{\bf\Delta}}({\bf
p})=D_q^{1/2}(R_{L_{\bf\Delta}}^W)D_c^{1/2}(R_{L_{
\bf\Delta}}^W)\Psi_{F\,{\bf 0}}({\bf p}),
\end{equation}
where $R^W$ is the Wigner rotation, $L_{\bf\Delta}$ is the Lorentz boost
from the meson rest frame to a moving one, and   
the rotation matrix $D^{1/2}(R)$ in spinor representation is given by
\begin{equation}\label{d12}
{1 \ \ \,0\choose 0 \ \ \,1}D^{1/2}_{q,c}(R^W_{L_{\bf\Delta}})=
S^{-1}({\bf p}_{q,c})S({\bf\Delta})S({\bf p}),
\end{equation}
where
$$
S({\bf p})=\sqrt{\frac{\epsilon(p)+m}{2m}}\left(1+\frac{\bm{\alpha}{\bf p}}
{\epsilon(p)+m}\right)
$$
is the usual Lorentz transformation matrix of the four-spinor.

The expressions for the matrix elements of the $B_c$ decays to the $P$-wave
$B_{sJ}$ and $B_J$ mesons, governed by the $c$ quark decays, can be
obtained from the above expressions by the interchange of the $b$ and $c$
quarks and for the final active quarks $q=s,d$. 

\section{Form factors of the semileptonic $B_c$ decays to the 
  orbitally excited heavy mesons}
\label{ffr}

The matrix elements of the weak current $J^W_\mu=\bar
b\gamma_\mu(1-\gamma_5)q$ or $\bar
c\gamma_\mu(1-\gamma_5)q$  for $B_c$ decays to orbitally
excited scalar light mesons ($S$) can be parametrized by two invariant
form factors
\begin{eqnarray}
  \label{eq:sff1}
\langle S(p_F)|\bar q \gamma^\mu b|B_c(p_{B_c})\rangle
  &=&0,\cr\cr
  \langle S(p_F)|\bar q \gamma^\mu\gamma_5 b|B_c(p_{B_c})\rangle
  &=&f_+(q^2)\left(p_{B_c}^\mu+ p_F^\mu\right)+
  f_-(q^2)\left(p_{B_c}^\mu- p_F^\mu\right),
\end{eqnarray}
where $q=p_{B_c}-p_F$, $M_S$ is the scalar meson mass.

The matrix elements of the weak current for $B_c$ decays to axial
vector mesons ($AV$)
can be expressed in terms of four invariant form factors
\begin{eqnarray}
  \label{eq:avff1}
  \langle A(p_F)|\bar q \gamma^\mu b|B_c(p_{B_c})\rangle&=&
  (M_{B_c}+M_A)h_{V_1}(q^2)\epsilon^{*\mu}
  +[h_{V_2}(q^2)p_{B_c}^\mu+h_{V_3}(q^2)p_F^\mu]\frac{\epsilon^*\cdot q}{M_{B_c}} ,\qquad\\\cr
\label{eq:avff2}
\langle A(p_F)|\bar q \gamma^\mu\gamma_5 b|B_c(p_{B_c})\rangle&=&
\frac{2ih_A(q^2)}{M_{B_c}+M_A} \epsilon^{\mu\nu\rho\sigma}\epsilon^*_\nu
  p_{B_c\rho} p_{F\sigma},  
\end{eqnarray}
where $M_A$ and $\epsilon^\mu$ are the mass and polarization vector of 
the axial vector meson.

The matrix elements of the weak current for $B_c$ decays to tensor
mesons ($T$)
can be decomposed in four Lorentz-invariant structures
\begin{eqnarray}
  \label{eq:tff1}
  \langle T(p_F)|\bar q \gamma^\mu b|B_c(p_{B_c})\rangle&=&
\frac{2it_V(q^2)}{M_{B_c}+M_T} \epsilon^{\mu\nu\rho\sigma}\epsilon^*_{\nu\alpha}
\frac{p_{B_c}^\alpha}{M_{B_c}}  p_{B_c\rho} p_{F\sigma},\\\cr
\label{eq:tff2}
\langle T(p_F)|\bar q \gamma^\mu\gamma_5 b|B_c(p_{B_c})\rangle&=&
(M_{B_c}+M_T)t_{A_1}(q^2)\epsilon^{*\mu\alpha}\frac{p_{B\alpha}}{M_{B_c}}\cr\cr
&&  +[t_{A_2}(q^2)p_{B_c}^\mu+t_{A_3}(q^2)p_F^\mu]\epsilon^*_{\alpha\beta}
\frac{p_{B_c}^\alpha p_{B_c}^\beta}{M_{B_c}^2} , 
\end{eqnarray}
where $M_T$ and $\epsilon^{\mu\nu}$ are the mass and polarization tensor of 
the tensor meson.

We previously studied the form factors parametrizing the matrix elements of
vector and axial vector charged and neutral weak currents for $B_c\to
\eta_c(J/\psi)$, $B_c\to D^{(*)}$ \cite{bcjpsi}, $B_c\to
B_s^{(*)}(B^{(*)})$ \cite{bcbs} and $B_c\to D_s^{(*)}$
 transitions in the framework of our model.  Now we apply the same
 approach, described in detail in 
Refs.~\cite{bcjpsi,bcbs,bdecays}, for the calculation of the form
factors for $B_c$ decays to the orbitally excited heavy
mesons. Namely, we calculate exactly the 
contribution of the leading vertex function $\Gamma^{(1)}$ 
(\ref{gamma1}) to the transition matrix element of the weak
current (\ref{mxet}) using the $\delta$-function.  For the evaluation of
the subleading contribution $\Gamma^{(2)}$ for the $B_c\to
\chi_J(h_c)$ and $B_c\to D_J$ transitions, governed by
$b\to c,u$ transitions,  we use expansions in inverse powers of the
heavy $b$-quark mass from  the initial $B_c$ meson and large recoil
energy of the final heavy meson. Note that the latter
contributions turn out to be rather small numerically. Therefore we
obtain reliable expressions for the form factors in the whole
accessible kinimatical range. It is important to emphasize that doing
these calculations we consistently take into account all relativistic
corrections including boosts of the meson wave functions from the rest reference
frame to the moving ones, given
by Eq.~(\ref{wig}).   The obtained expressions for the decay
form factors are given in Appendix (to simplify
these expressions the long-range anomalous chromomagnetic quark moment
was explicitly set as $\kappa=-1$). In
the limits of infinitely heavy quark mass and large energy of the final meson, the form
factors in our model satisfy all heavy quark symmetry relations
\cite{clopr,ffhm}.     

As a result, we get the following expressions for the $B_c$ decay
form factors:

(a) $B_c\to S$ transitions ($S=\chi_{c0},D_0$) 
\begin{equation}
  \label{eq:f+}
  f_\pm(q^2)=f_\pm^{(1)}(q^2)+\varepsilon f_\pm^{S(2)}(q^2)
+(1-\varepsilon) f_\pm^{V(2)}(q^2),
\end{equation}

(b) $B_c\to AV$ transition ($AV=\chi_{c1},D_1(^3P_1)$)
\begin{eqnarray}
  \label{eq:hV}
  h_{V_i}(q^2)&=&h_{V_i}^{(1)}(q^2)+\varepsilon h_{V_i}^{S(2)}(q^2)
+(1-\varepsilon) h_{V_i}^{V(2)}(q^2),\qquad (i=1,2,3),\cr\cr
 h_A(q^2)&=&h_A^{(1)}(q^2)+\varepsilon h_A^{S(2)}(q^2)
+(1-\varepsilon) h_A^{V(2)}(q^2),
\end{eqnarray}

(c) $B_c\to AV'$ transition\footnote{The corresponding decay matrix elements are
  defined by Eqs.~(\ref{eq:avff1}) and (\ref{eq:avff2}) with the
  replacement of form factors $h_i(q^2)$ by $g_i(q^2)$.} ($AV'=h_{c},D_1(^1P_1)$)
\begin{eqnarray}
  \label{eq:gV}
  g_{V_i}(q^2)&=&g_{V_i}^{(1)}(q^2)+\varepsilon g_{V_i}^{S(2)}(q^2)
+(1-\varepsilon) g_{V_i}^{V(2)}(q^2),\qquad (i=1,2,3),\cr\cr
 g_A(q^2)&=&g_A^{(1)}(q^2)+\varepsilon g_A^{S(2)}(q^2)
+(1-\varepsilon) g_A^{V(2)}(q^2),
\end{eqnarray}

(d) $B_c\to T$ transition ($T=\chi_{c2},D_2^*$)
\begin{eqnarray}
  \label{eq:tV}
 t_V(q^2)&=&t_V^{(1)}(q^2)+\varepsilon t_V^{S(2)}(q^2)
+(1-\varepsilon) t_V^{V(2)}(q^2),\cr\cr
  t_{A_i}(q^2)&=&t_{A_i}^{(1)}(q^2)+\varepsilon t_{A_i}^{S(2)}(q^2)
+(1-\varepsilon) t_{A_i}^{V(2)}(q^2),\qquad (i=1,2,3),
\end{eqnarray}
where $f_{\pm}^{(1)}$, $f_{\pm}^{S,V(2)}$, $h_{V_i}^{(1)}$,
$h_{V_i}^{S,V(2)}$, $h_A^{(1)}$, $h_A^{S,V(2)}$,  $g_{V_i}^{(1)}$,
$g_{V_i}^{S,V(2)}$, $g_A^{(1)}$, $g_A^{S,V(2)}$, $t_V^{(1)}$,
$t_V^{S,V(2)}$, $t_{A_i}^{(1)}$, and
$t_{A_i}^{S,V(2)}$ are given in Appendix.
The superscripts ``(1)" and ``(2)" correspond to Figs.~\ref{d1} and
\ref{d2}, $S$ and
$V$ correspond to the scalar and vector confining potentials of the $q\bar q$-interaction.
The mixing parameter of scalar and vector confining potentials
$\varepsilon$ is fixed to be $-1$  in our model.

 In the case of $B_c$ decays to $P$-wave $B_s$ and
$B$ mesons, governed by the $c\to s,d$ transitions, the accessible
kinematical range is significantly smaller (by almost a factor of 4)
than the one for the decays to the $S$-wave  $B_s$ and $B$ mesons. Our
previous investigation \cite{bcbs} of the latter decays had
shown that intermediate negative-energy states, leading to the
subleading term $\Gamma^{(2)}$, give almost negligible
contributions to decay form factors (see Fig.~3 of
Ref.~\cite{bcbs}). Therefore such contributions can be safely
neglected in the present analysis. Thus, for calculations of the form
factors of the $B_c\to B_{sJ}$ and $B_c\to B_{J}$ weak transitions we
use the leading order expressions $f_i^{(1)}$, $h_i^{(1)}$,
$g_i^{(1)}$ and $t_i^{(1)}$, given in Appendix, where the $b$ and $c$
quarks are interchanged.

 For numerical calculations of the form factors we use the
quasipotential wave functions of the 
$B_c$ meson and orbitally excited charmonium and $D$, $B_s$, $B$ mesons
obtained in their mass spectra 
calculations \cite{mass,hlm}. Our results for the masses of these
mesons are in good agreement with available experimental data
\cite{pdg}. Therefore we use the experimental values for the masses of
well-established states and our model predictions for all other masses
in the numerical calculations. 

\begin{figure}
  \centering
  \includegraphics[width=7.5cm]{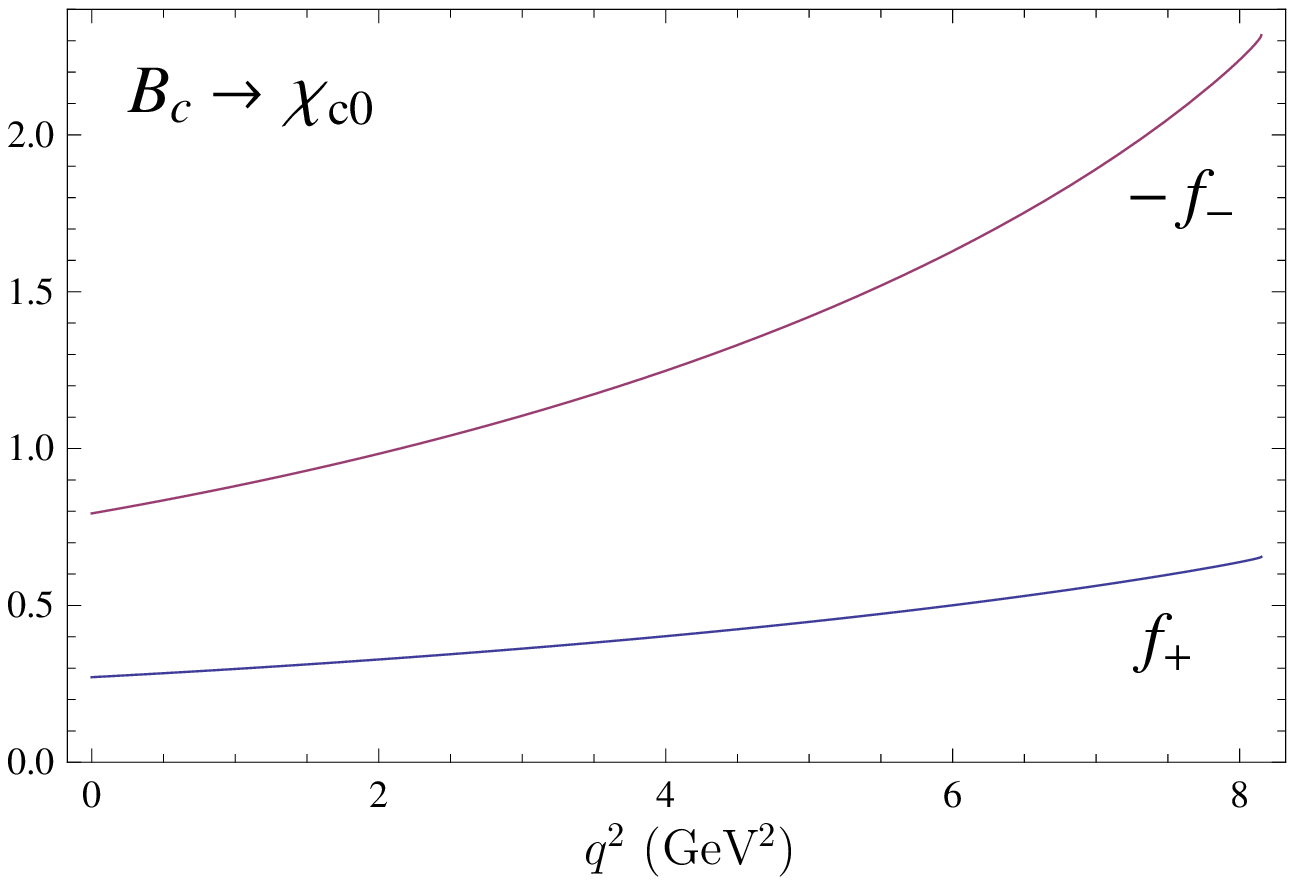} \ \ \ \
\  
\includegraphics[width=7.5cm]{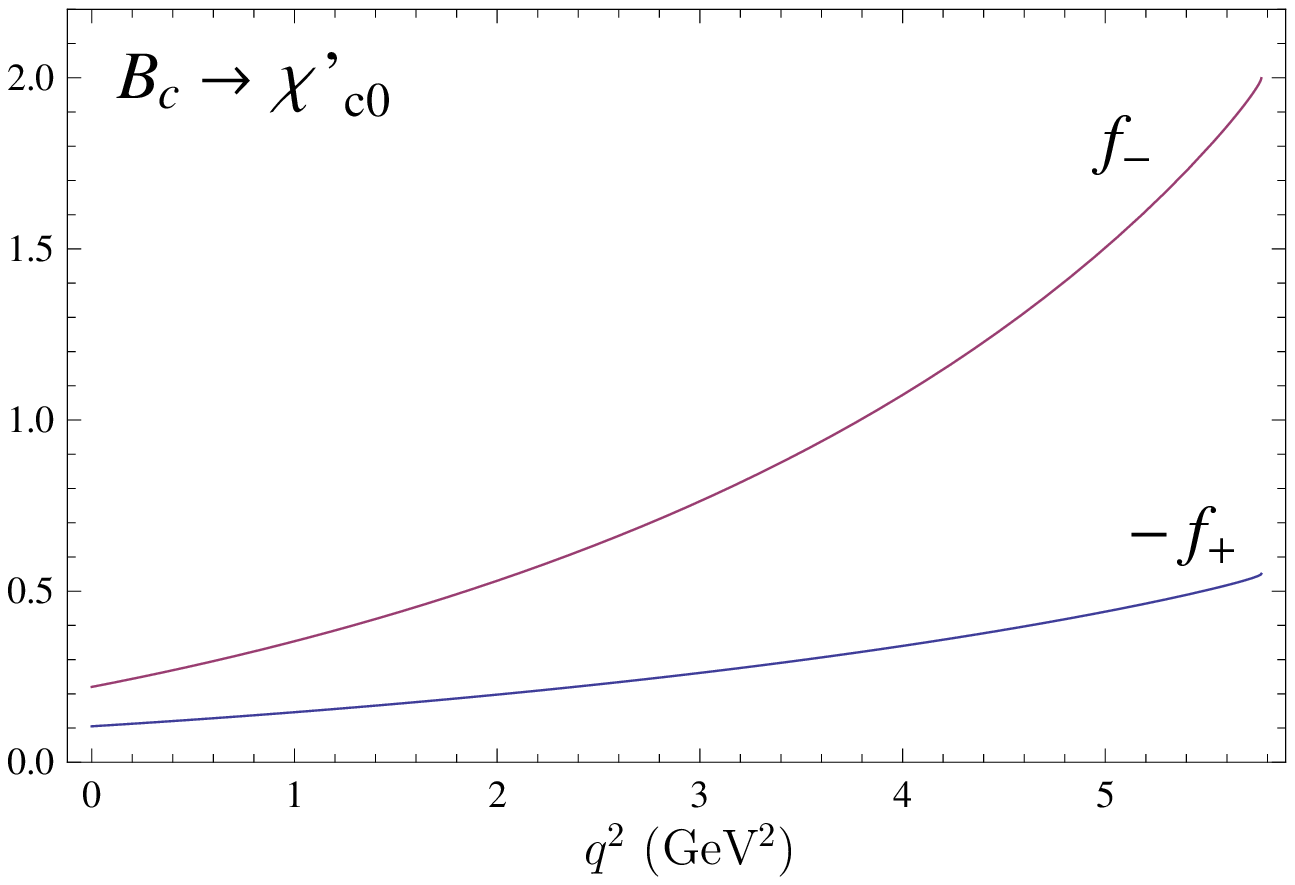}

\vspace*{0.5cm}

  \includegraphics[width=7.5cm]{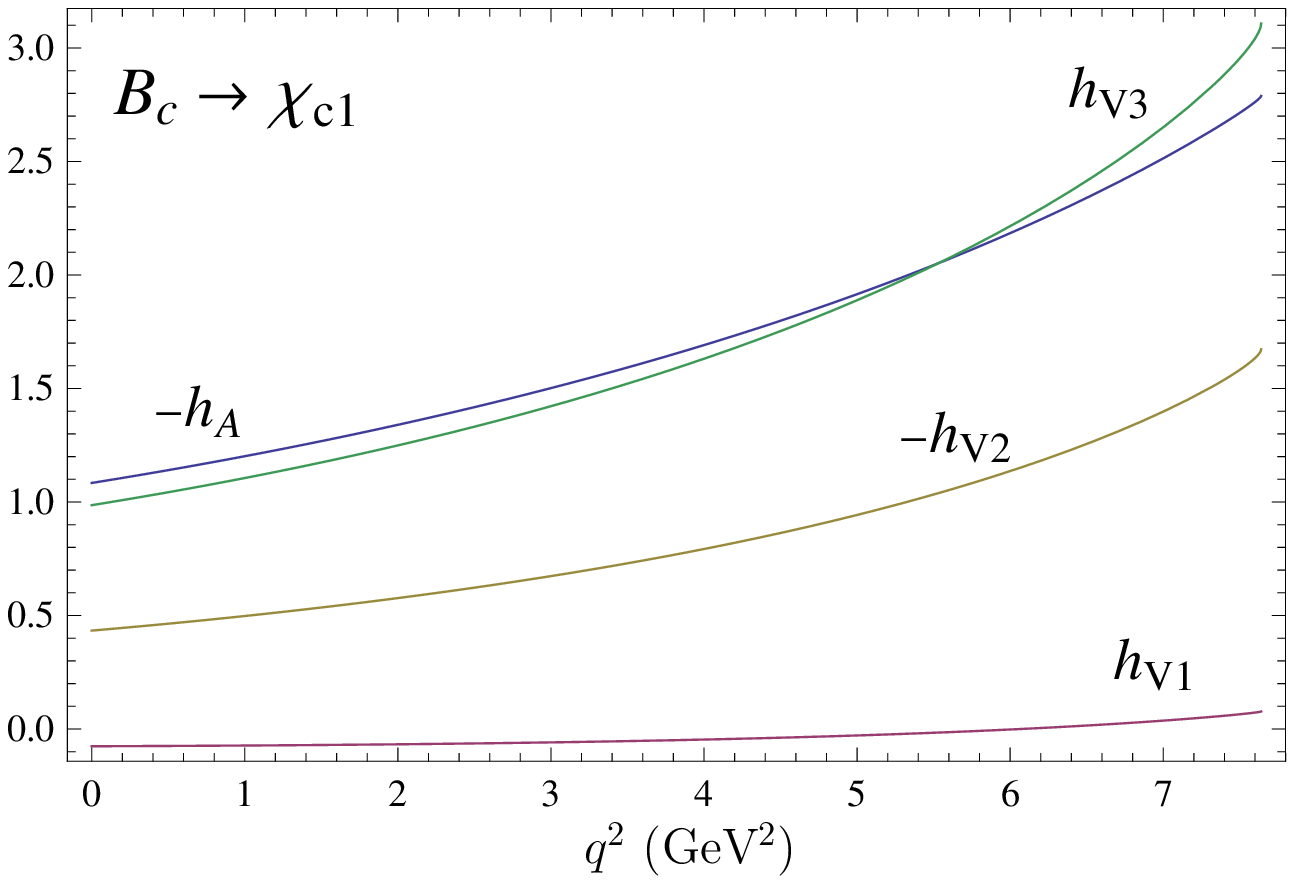} \ \ \ \
\  
\includegraphics[width=7.5cm]{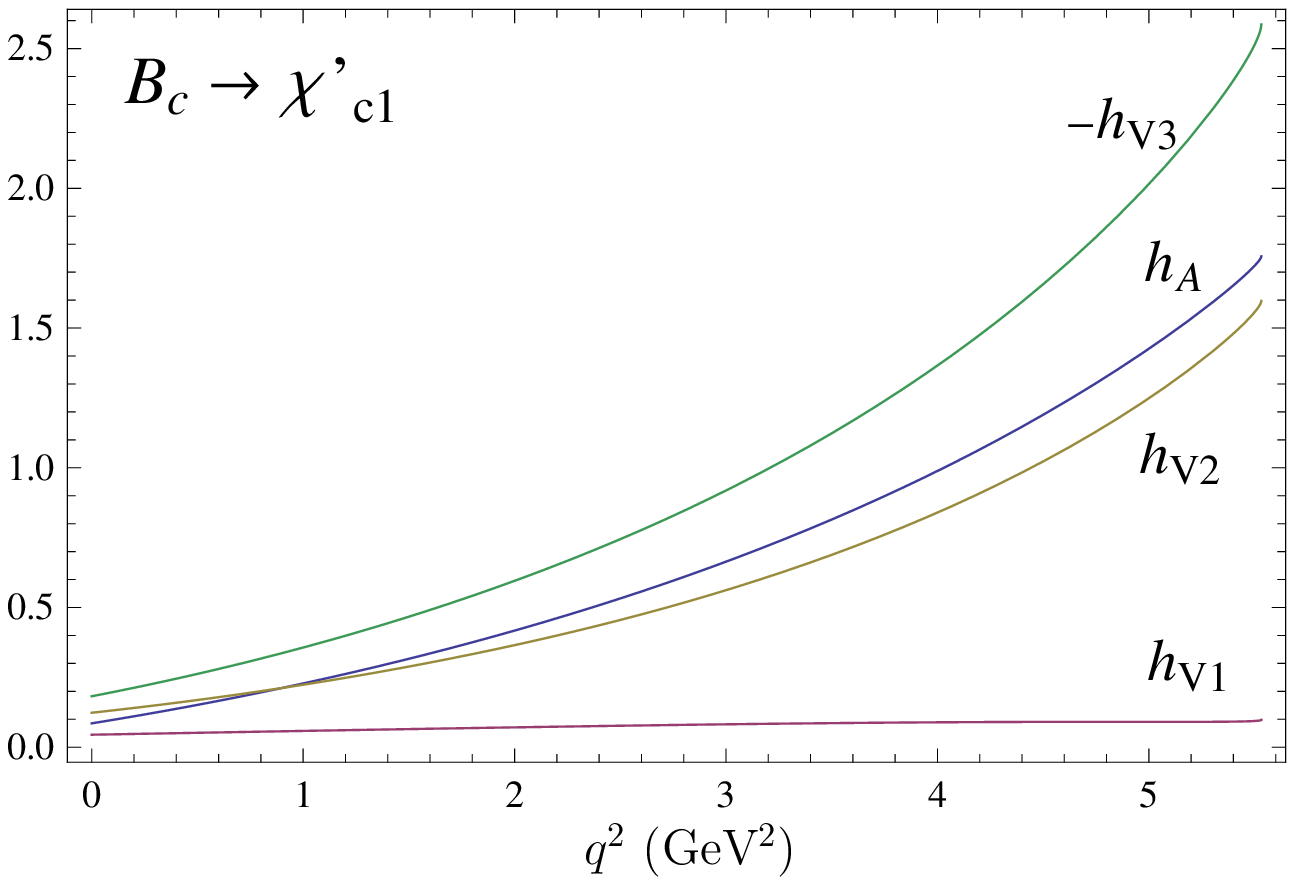}

\vspace*{0.5cm}
  \includegraphics[width=7.5cm]{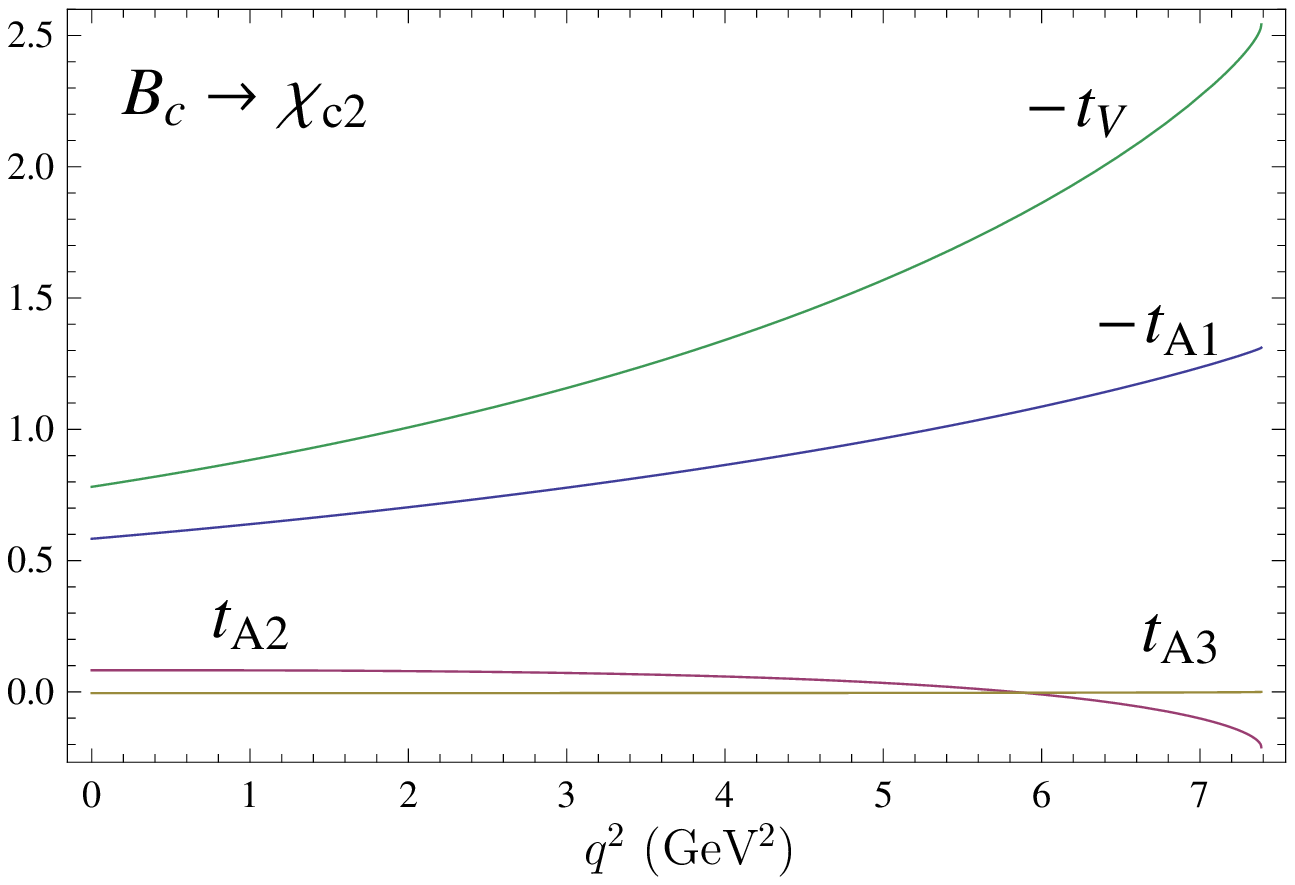} \ \ \ \
\  
\includegraphics[width=7.5cm]{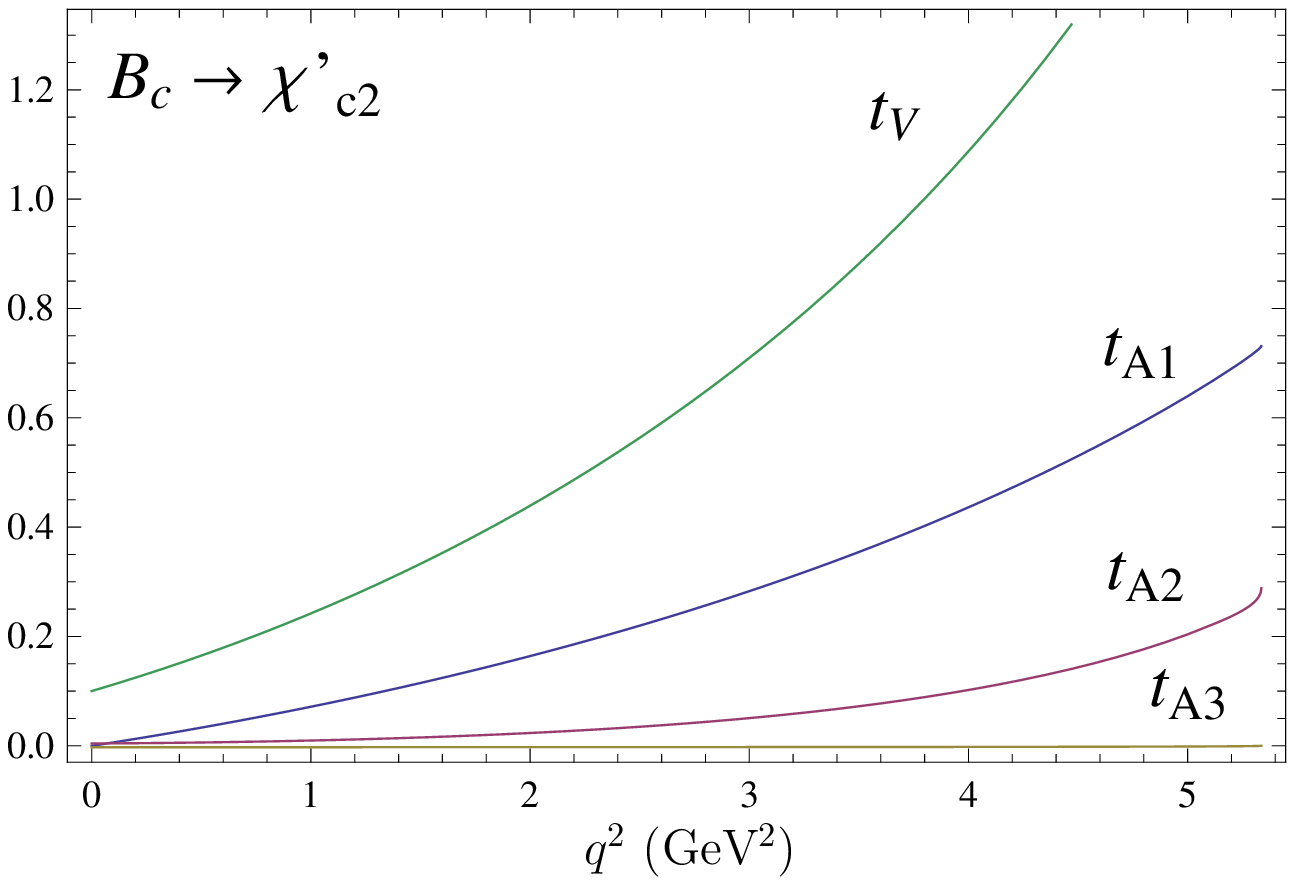}

\vspace*{0.5cm}

  \includegraphics[width=7.5cm]{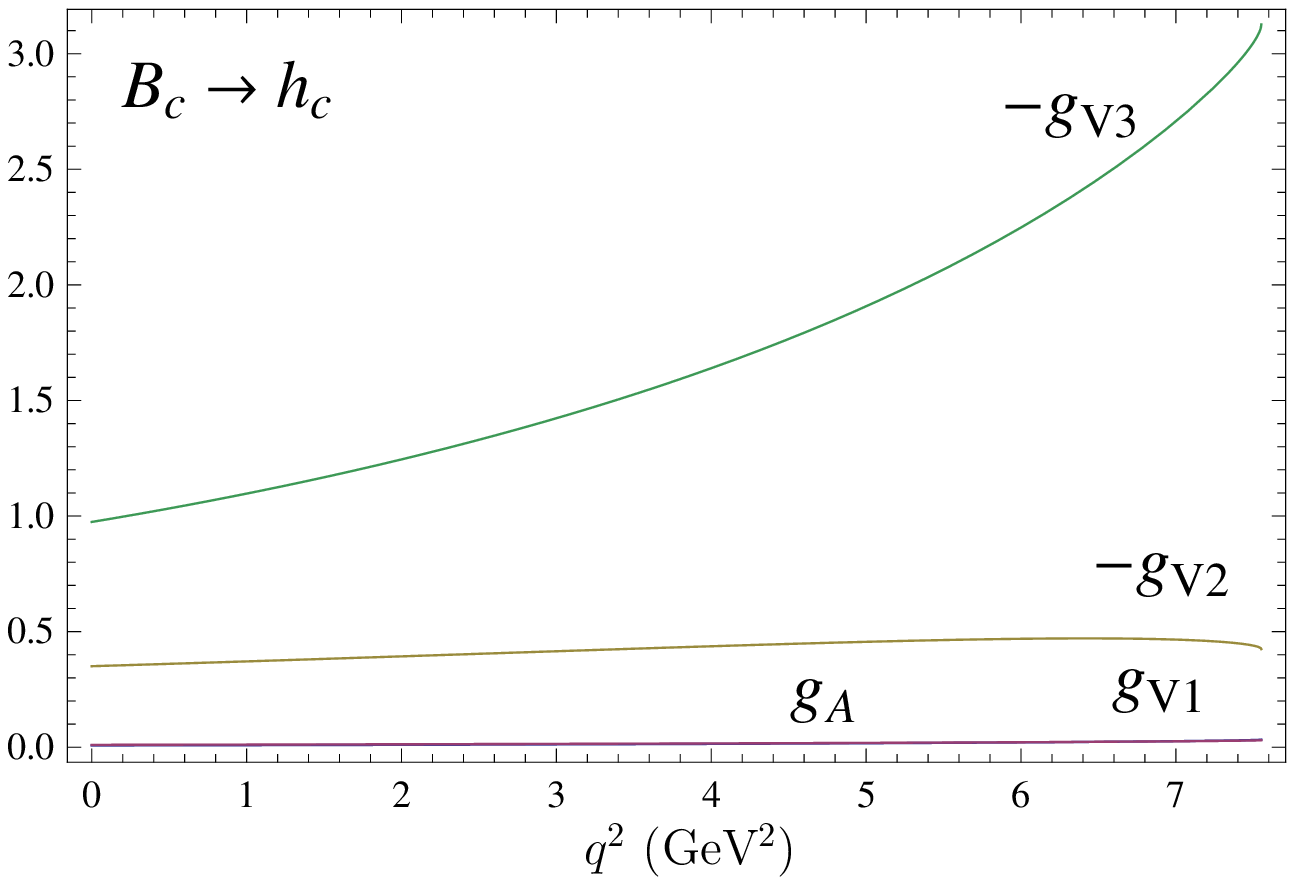} \ \ \ \
\  
\includegraphics[width=7.5cm]{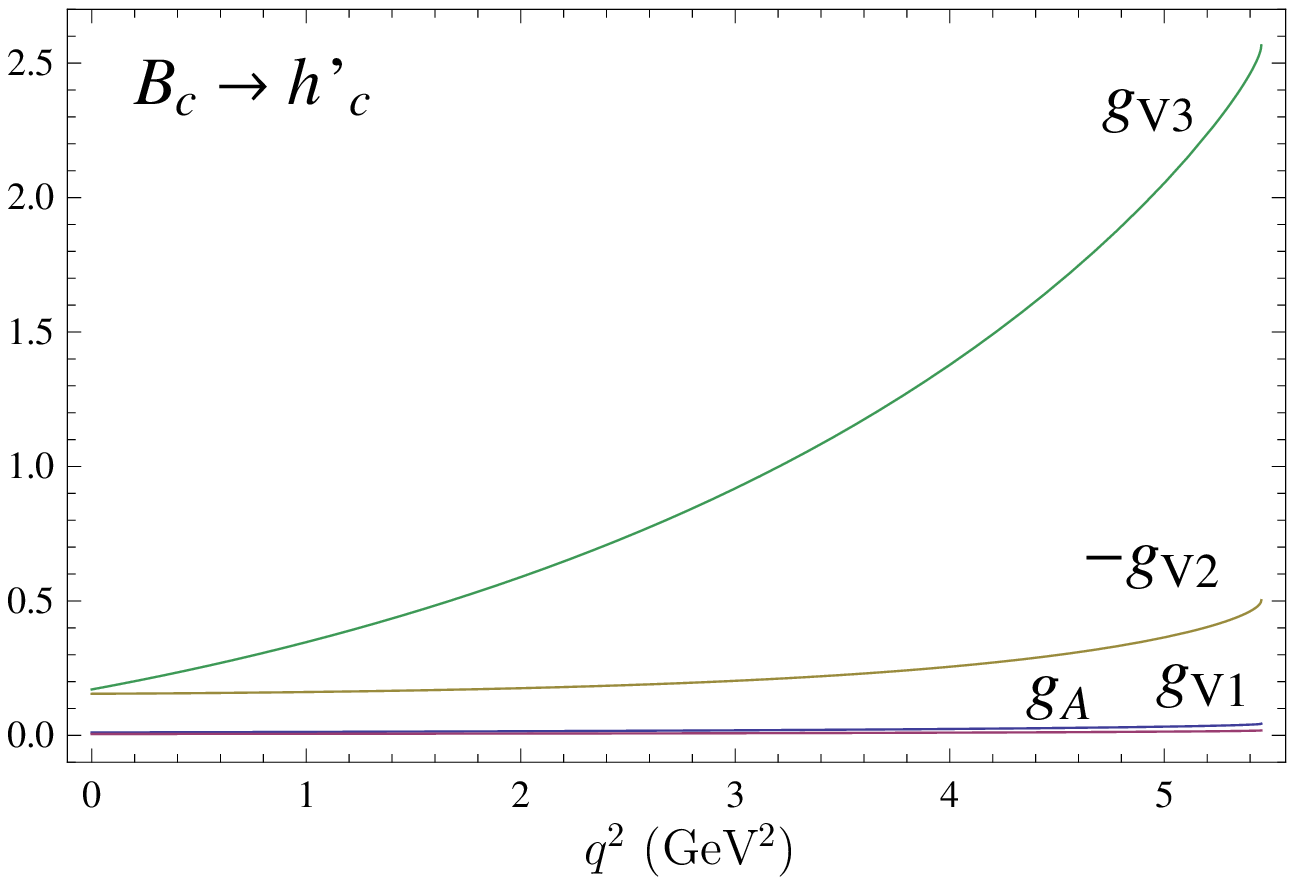}
  \caption{Form factors of the $B_c$ decays to the $1P$- and $2P$ -wave charmonium states.}
  \label{fig:ffbc}
\end{figure}

In Fig.~\ref{fig:ffbc} we plot form factors of the $B_c$ weak transitions
to the $1P$ ($\chi_{cJ},h_c$) and $2P$ ($\chi'_{cJ},h'_c$) -wave
charmonium states as an example. The 
remaining plots for the $B_c$ weak form factors to the $P$-wave $D_J$, $B_{sJ}$ and
$B_J$ mesons have an analogous behaviour and are not shown here.

\section{Semileptonic $B_c$ decays to orbitally excited heavy mesons}
\label{sec:sd}

The differential decay rate for the $B_c$ meson decay to $P$-wave heavy mesons reads \cite{iks}
\begin{equation}
  \label{eq:dgamma}
  \frac{d\Gamma(B_c\to F(S,AV,T)l\bar\nu)}{dq^2}=\frac{G_F^2}{(2\pi)^3}
  |V_{bf}|^2
  \frac{\lambda^{1/2}(q^2-m_l^2)^2}{24M_{B_c}^3q^2} 
  \Biggl[H H^{\dag}\left(1+\frac{m_l^2}{2q^2}\right)  +\frac{3m_l^2}{2q^2} H_tH^{\dag}_t\Biggr],
\end{equation}
where $G_F$ is the Fermi constant, $V_{ij}$ are the Cabbibo-Kobayashi-Maskawa (CKM) matrix elements, $\lambda\equiv
\lambda(M_{B_c}^2,M_F^2,q^2)=M_{B_c}^4+M_F^4+q^4-2(M_{B_c}^2M_F^2+M_F^2q^2+M_{B_c}^2q^2)$,
$m_l$ is the lepton mass and
\begin{equation}
  \label{eq:hh}
  H H^{\dag}\equiv H_+H^{\dag}_++H_-H^{\dag}_-+H_0H^{\dag}_0.
\end{equation}
Helicity components of the hadronic tensor are expressed through the
invariant form factors.

(a) $B_c\to S(^3P_0)$ transition
\begin{eqnarray}
  \label{eq:has}
  H_\pm&=&0,\cr
H_0&=&\frac{\lambda^{1/2}}{\sqrt{q^2}}f_+(q^2),\cr
H_t&=&\frac1{\sqrt{q^2}}[(M_{B_c}^2-M_S^2)f_+(q^2)+q^2f_-(q^2)].
\end{eqnarray}

(b) $B_c\to AV(^3P_1)$ transition
\begin{eqnarray}
  \label{eq:haav}
  H_\pm&=&(M_{B_c}+M_{AV})h_{V_1}(q^2) \pm\frac{\lambda^{1/2}}{M_{B_c}+M_{AV}}h_A,\cr
H_0&=&\frac{1}{2M_{AV}\sqrt{q^2}} \left\{(M_{B_c}+M_{AV})(M_{B_c}^2-M_{AV}^2-q^2)h_{V_1}(q^2)+\frac{\lambda}{2M_{B_c}}[h_{V_2}(q^2)+h_{V_3}(q^2)]\right\},\cr
H_t&=&\frac{\lambda^{1/2}}{2M_{AV}\sqrt{q^2}}
\Biggl\{(M_{B_c}+M_{AV})h_{V_1}(q^2)+\frac{M_{B_c}^2-M_{AV}^2}{2M_{B_c}}[h_{V_2}(q^2)+h_{V_3}(q^2)]\cr
&&+\frac{q^2}{2M_{B_c}}[h_{V_2}(q^2)-h_{V_3}(q^2)]\Biggr\}.
\end{eqnarray}

(c) $B_c\to AV'(^1P_1)$ transition\\
\nopagebreak
\indent $H_i$ are obtained from
expressions (\ref{eq:haav}) by replacement of form factors $h_i(q^2)$
by $g_i(q^2)$.

(d) $B_c\to T(^3P_2)$ transition
\begin{eqnarray}
  \label{eq:haat}
  H_\pm&=&\frac{\lambda^{1/2}}{2\sqrt{2}M_{B_c}M_{T}}\left[(M_{B_c}+M_{T})t_{A_1}(q^2) \pm\frac{\lambda^{1/2}}{M_{B_c}+M_{T}}t_V\right],\cr
H_0&=&\frac{\lambda^{1/2}}{2\sqrt{6}M_{B_c}M_{T}^2\sqrt{q^2}} \left\{(M_{B_c}+M_{T})(M_{B_c}^2-M_{T}^2-q^2)t_{A_1}(q^2)+\frac{\lambda}{2M_{B_c}}[t_{A_2}(q^2)+t_{A_3}(q^2)]\right\},\cr
H_t&=&\sqrt{\frac23}\frac{\lambda}{4M_{B_c}M_{T}^2\sqrt{q^2}}
\Biggl\{(M_{B_c}+M_{T})t_{A_1}(q^2)+\frac{M_{B_c}^2-M_{T}^2}{2M_{B_c}}[t_{A_2}(q^2)+t_{A_3}(q^2)]\cr
&&+\frac{q^2}{2M_{B_c}}[t_{A_2}(q^2)-t_{A_3}(q^2)]\Biggr\}.
\end{eqnarray}
Here the subscripts $\pm,0,t$ denote transverse, longitudinal and time helicity
components, respectively.

\begin{figure}
  \centering
 \includegraphics[width=8cm]{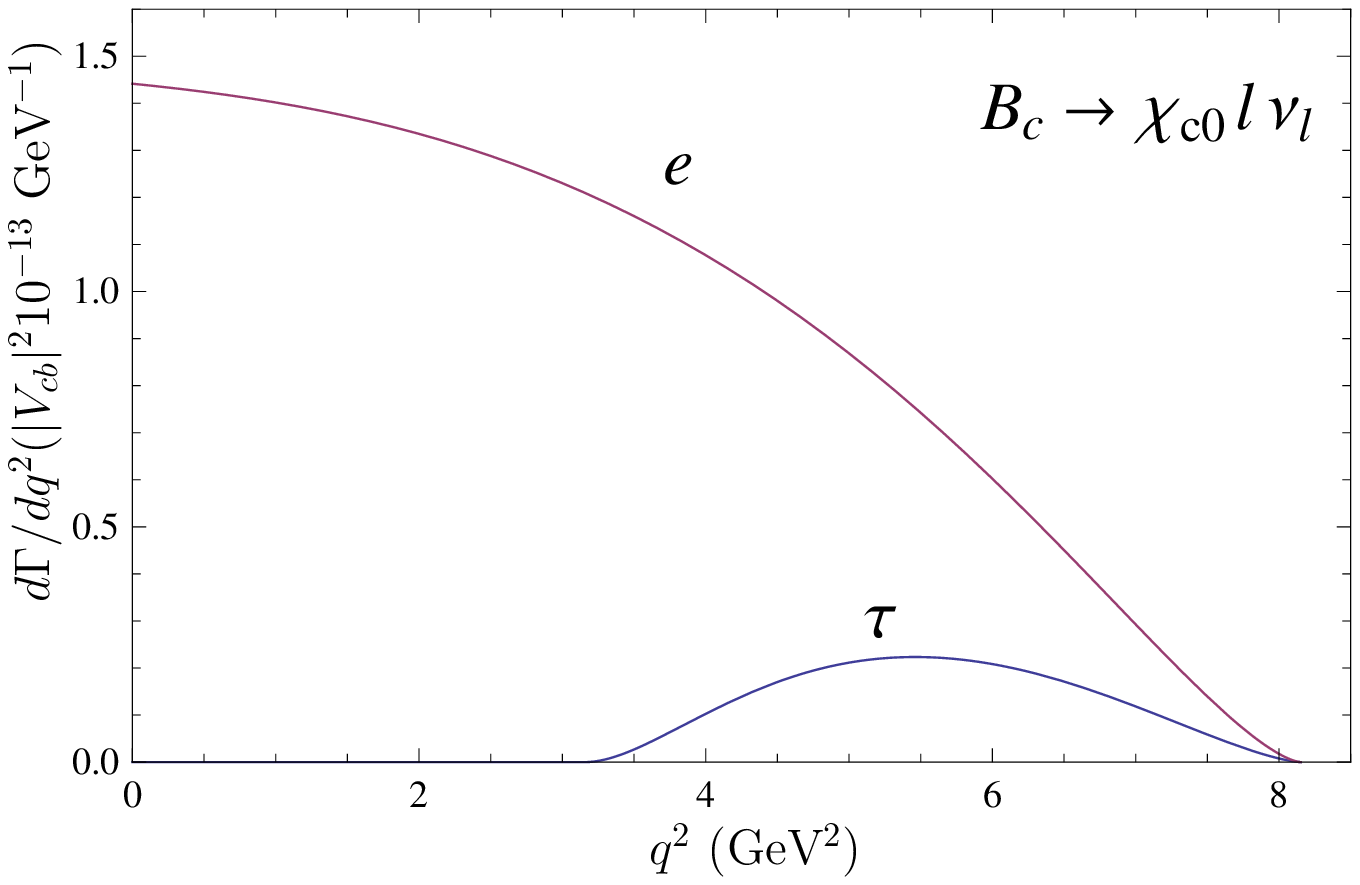}\ \
 \  \includegraphics[width=8cm]{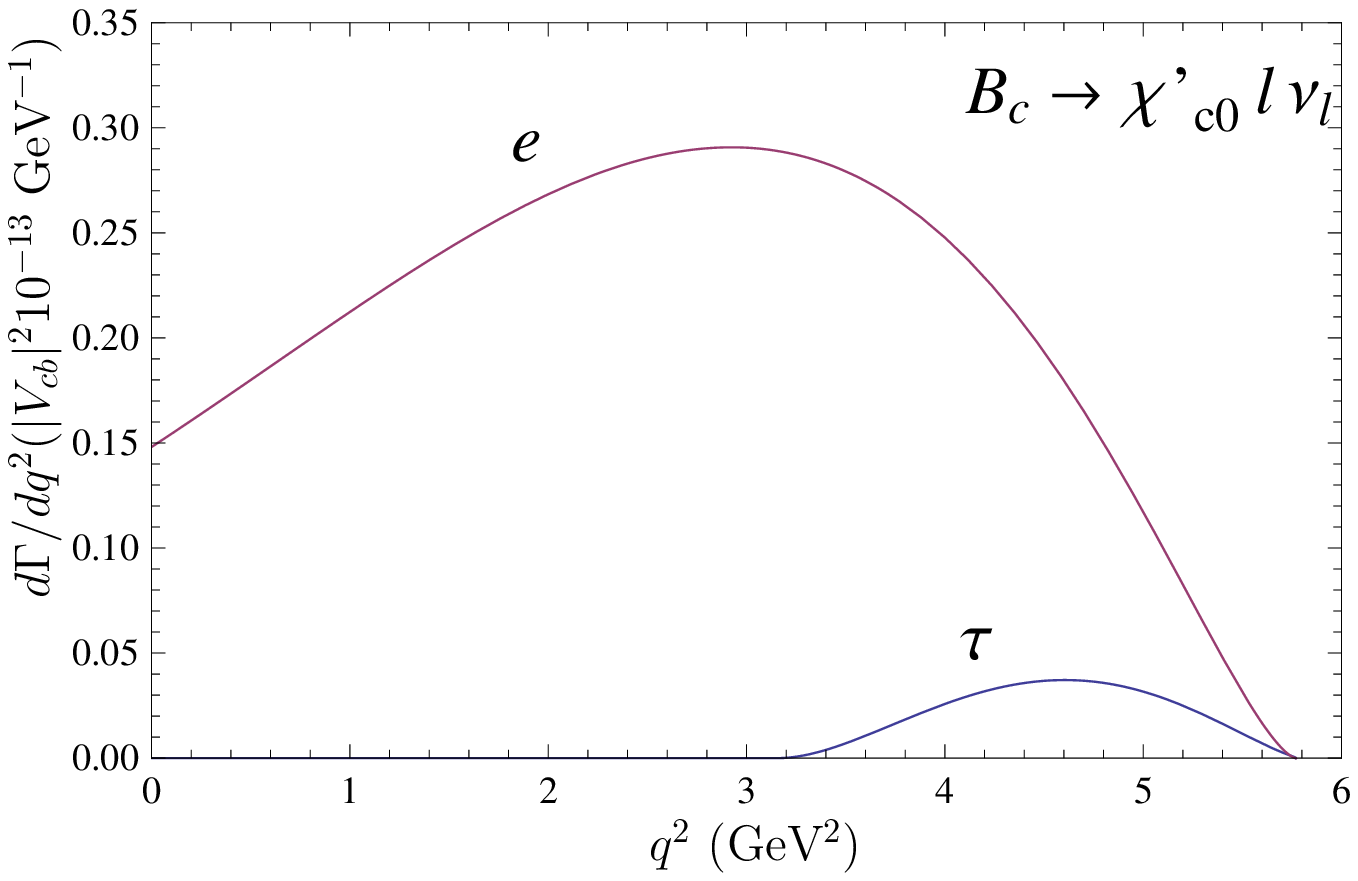}

\vspace*{0.5cm}
 \includegraphics[width=8cm]{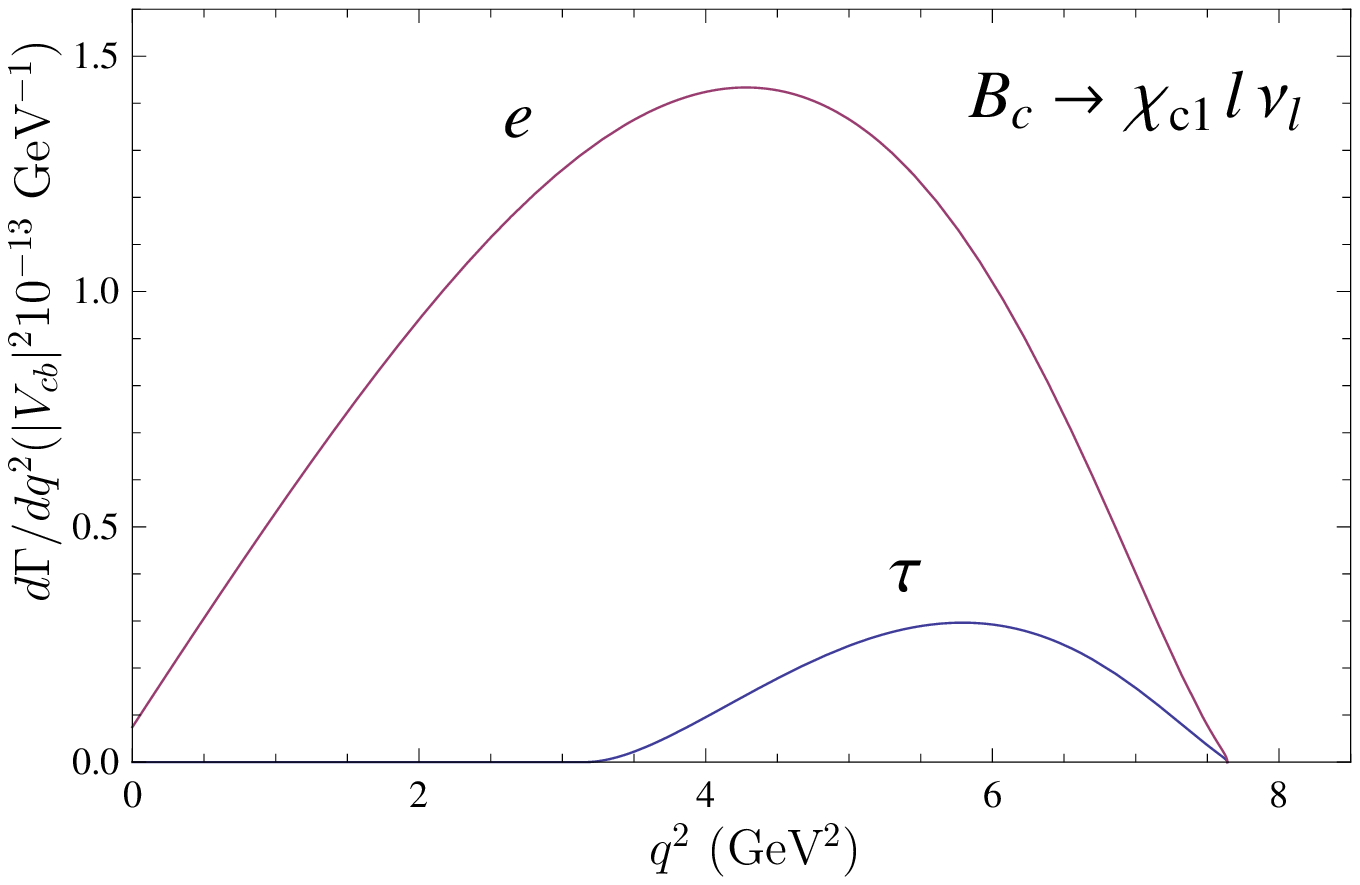}\ \ \  \includegraphics[width=8cm]{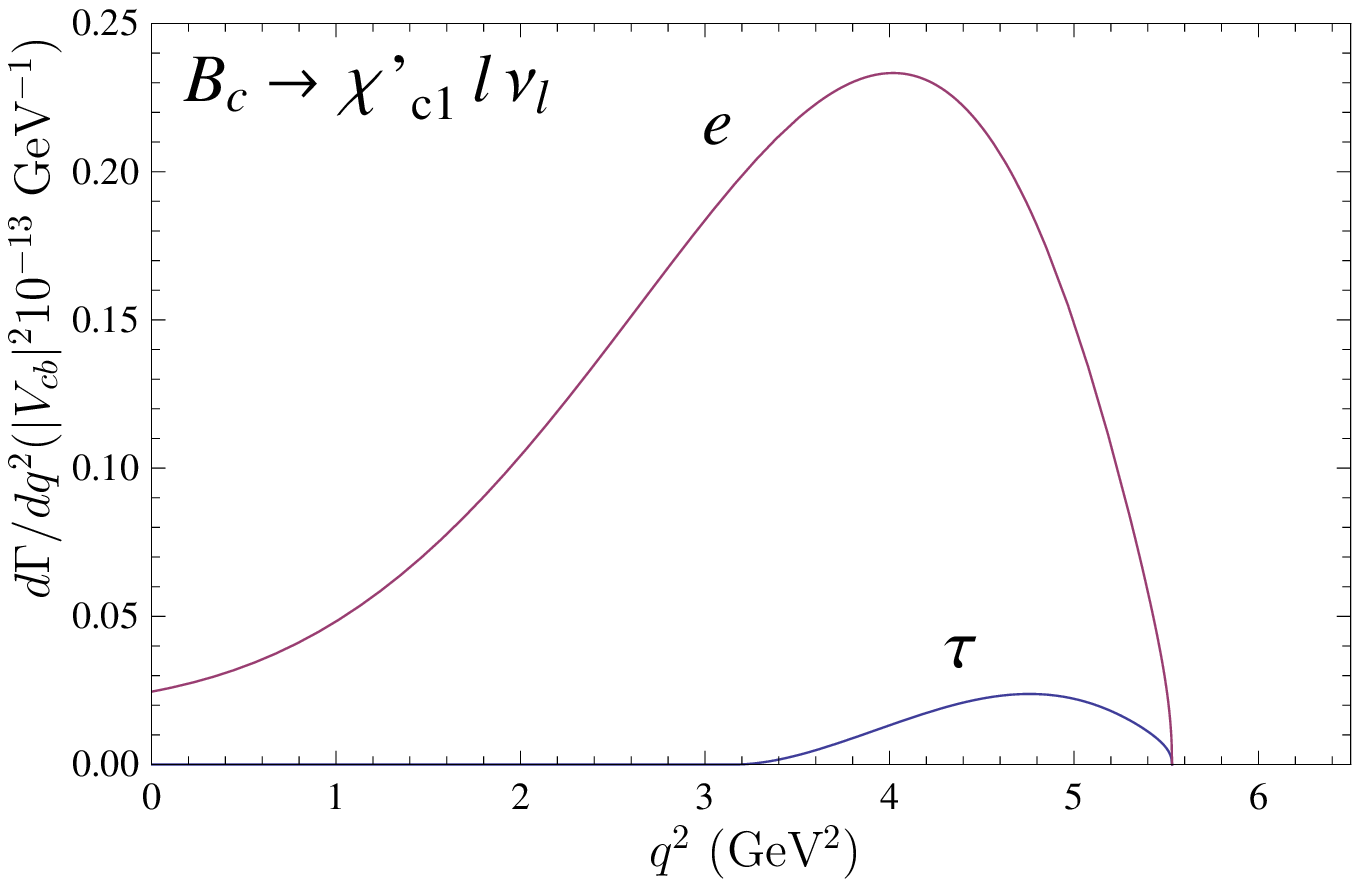}

\vspace*{0.5cm}
 \includegraphics[width=8cm]{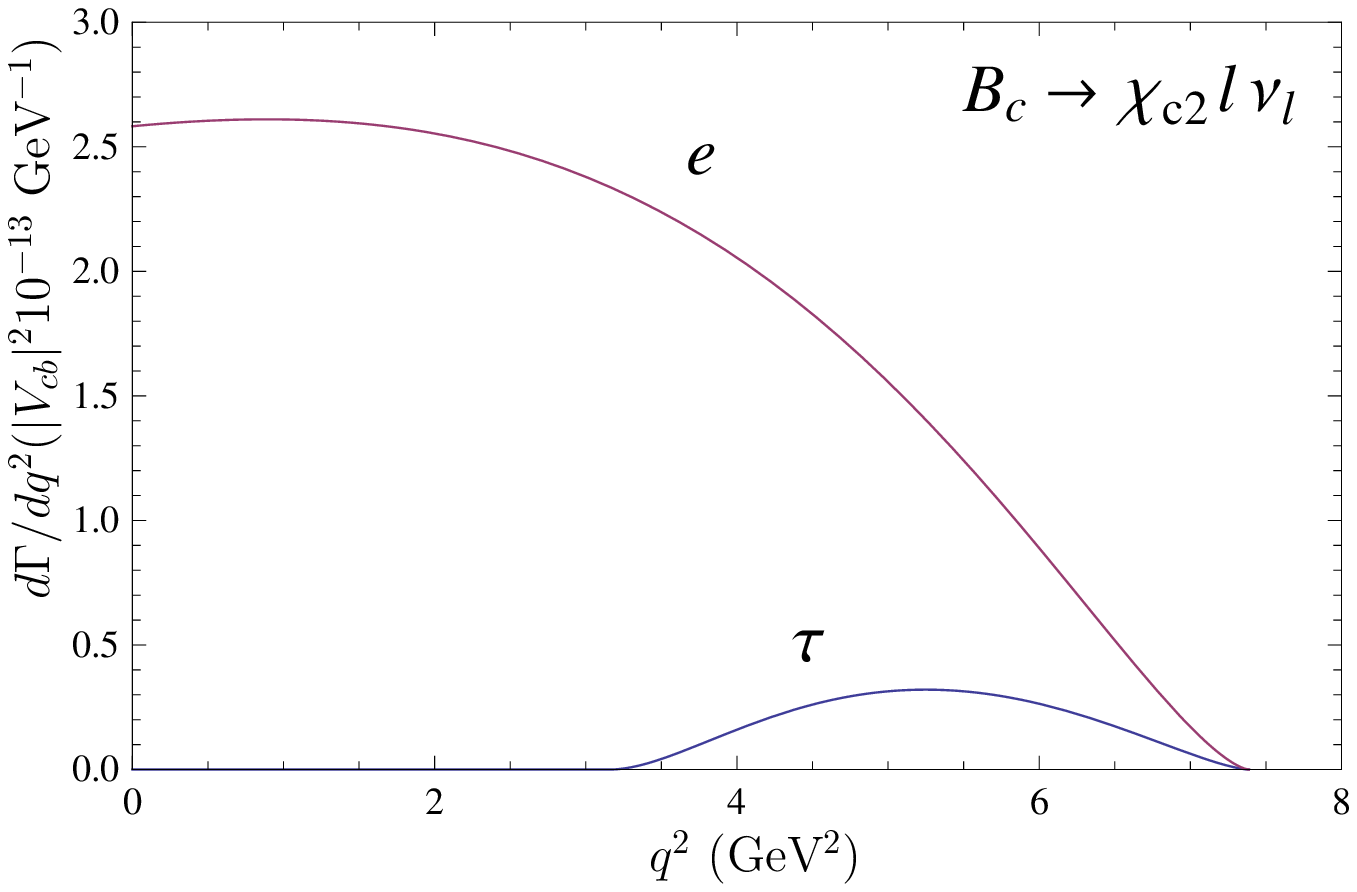}\ \
 \  \includegraphics[width=8cm]{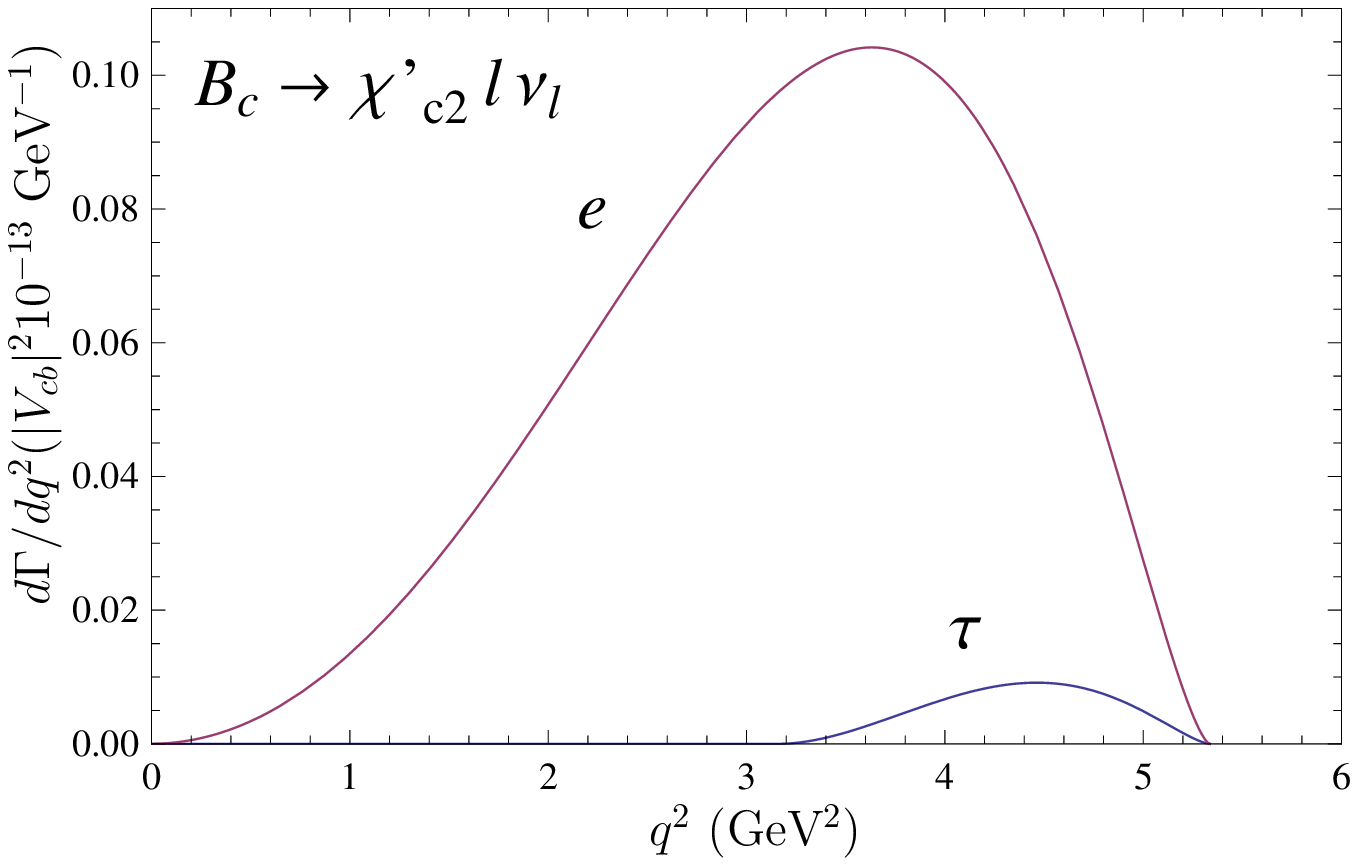}

\vspace*{0.5cm}
 \includegraphics[width=8cm]{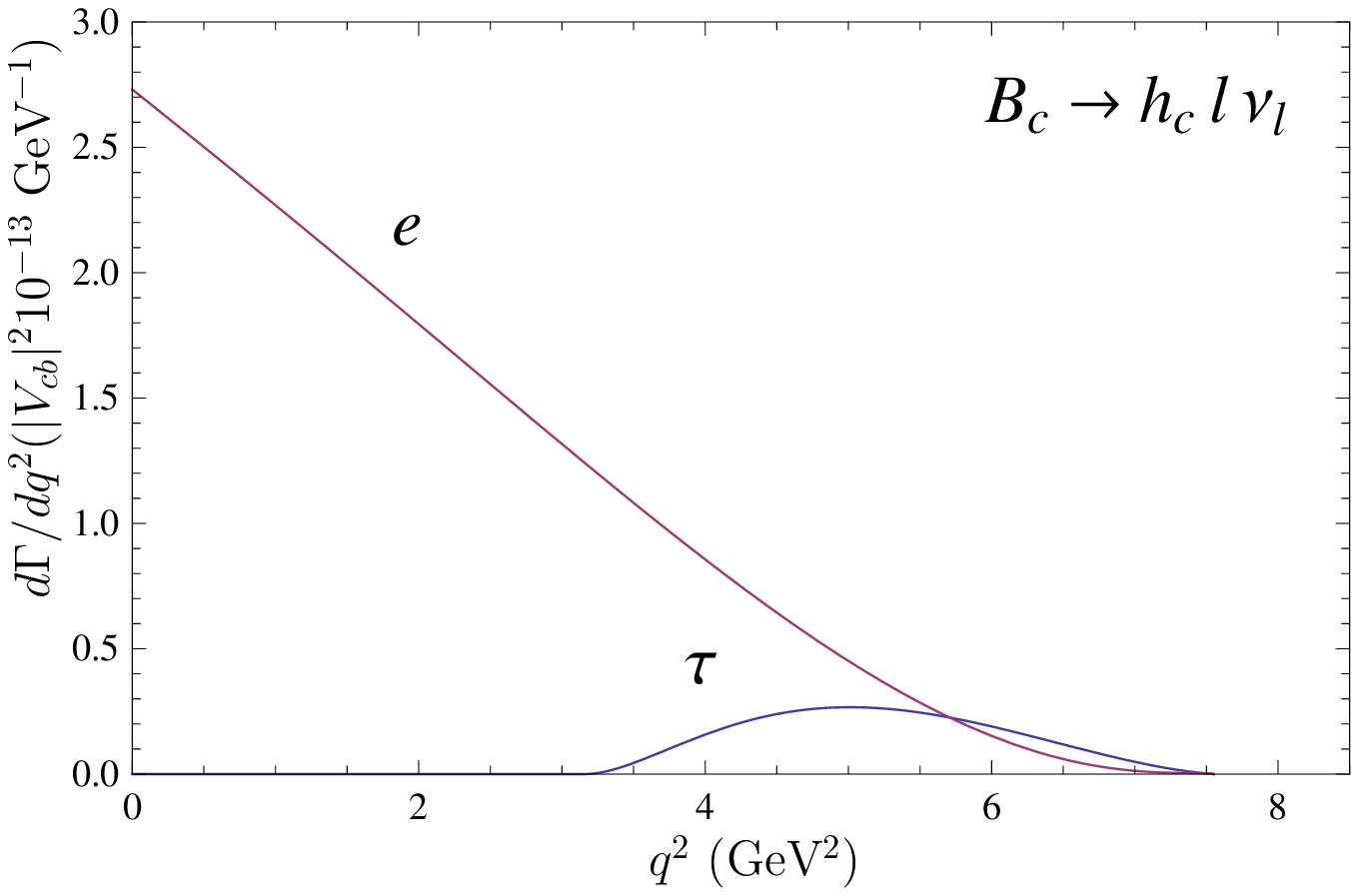}\ \ \  \includegraphics[width=8cm]{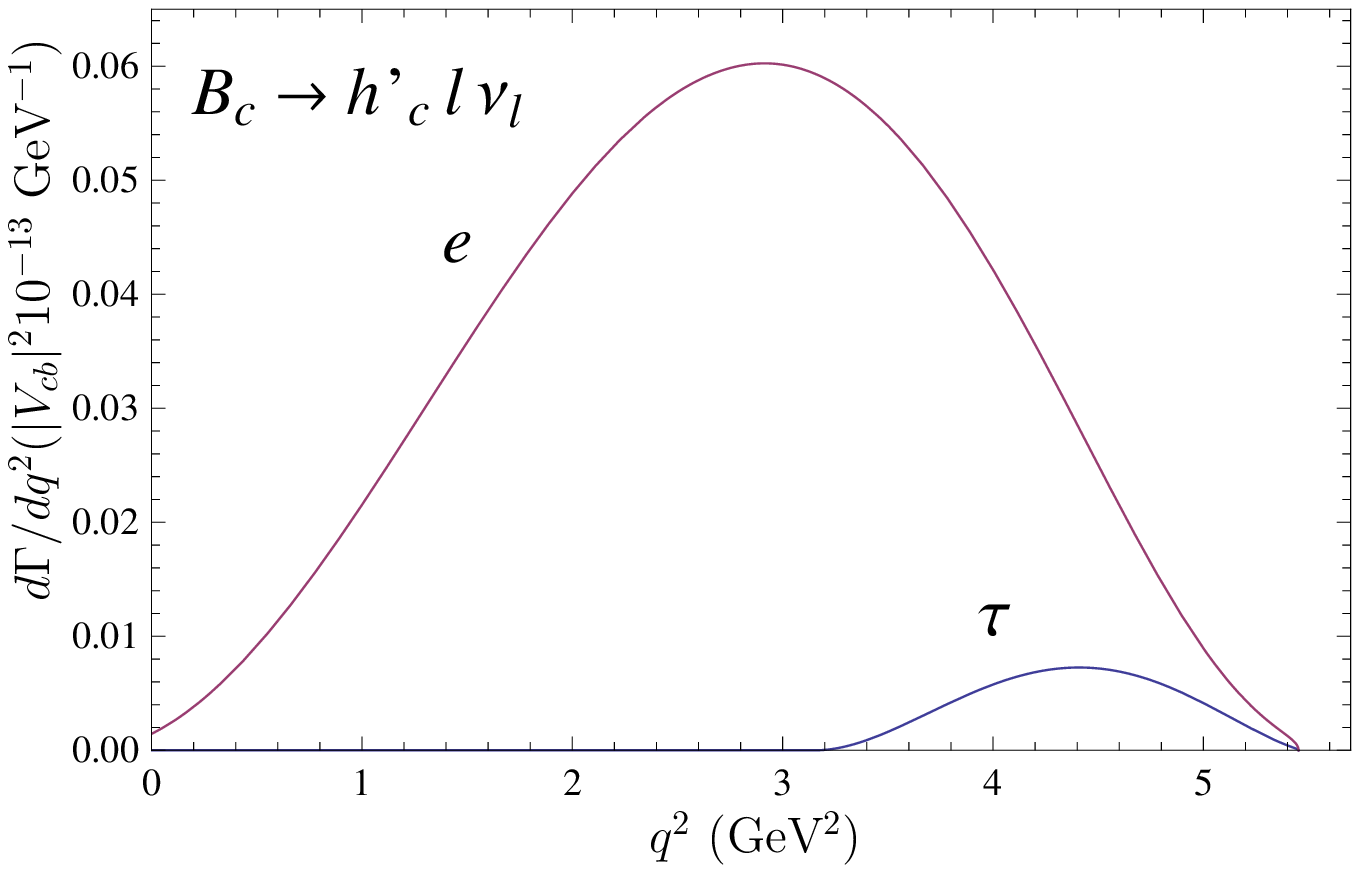}

  \caption{Predictions for the differential decay rates  of the $B_c$ 
    semileptonic decays to the $1P$- and $2P$-wave charmonium states. }
  \label{fig:brbc}
\end{figure}

Now we substitute the weak decay form factors calculated in the
previous section in the above expressions for decay
rates. The resulting differential distributions for $B_c$ decays to
the $1P$ ($\chi_J,h_c$) and $2P$ ($\chi'_J,h'_c$) charmonium states
are plotted in Fig.~\ref{fig:brbc}. The difference of the plot shapes
for the corresponding $1P$ and $2P$ charmonium states is the
consequence of their different nodal structure. We calculate the total rates of the
semileptonic $B_c$ decays  to the $P$-wave heavy mesons by integrating
the corresponding  differential decay rates over $q^2$. For
calculations we use the following values of the CKM matrix elements:
$|V_{cb}|=0.041$, $|V_{ub}|=0.0038$, $|V_{cs}|=0.974$, 
$|V_{cd}|=0.223$. Our predictions for the  rates of the
semileptonic $B_c$ decays to the $P$-wave charmonium states are
compared with the previous calculations \cite{iks,hnv,ccwz,asb,wl} in
Table~\ref{brbc}. The authors of Refs.~\cite{iks,hnv,ccwz} use
different types of relativistic quark models. Calculations in
Ref.~\cite{asb} are based on the three-point QCD sum rules, while
Ref.~\cite{wl} employs the light-cone QCD sum rules. We find
that significantly different theoretical approaches give values
for the $B_c\to\chi_J(h_c)l\nu$ decay rates 
consistent in the order of magnitude, while for the $B_c$ decays to
first radial excitations of the $P$-wave charmonium
($B_c\to\chi'_J(h'_c)l\nu$) our results are almost an order of
magnitude lower than the predictions of the light-cone QCD sum rules \cite{wl},
which are the only available ones at present.  The latter decays can play
an important role 
in studying charmonium states above the open charm production
threshold. Their observation at Tevatron and LHC can help to clarify
the nature of the new charmonium-like states. 
Our results for the rates of the CKM suppressed  semileptonic $B_c$ decays
to the $P$-wave $D$ mesons, governed by the weak $b\to u$ transitions,
are given in Table~\ref{brbcd}.  

\begin{table}
\caption{Comparison of our predictions for the  rates of the
  semileptonic $B_c$ decays 
 to the $P$-wave charmonium states with previous calculations (in
  $10^{-15}$ GeV). }
\label{brbc}
\begin{ruledtabular}
\begin{tabular}{ccccccc}
 Decay& our &\cite{iks} &\cite{hnv}&  \cite{ccwz}& \cite{asb}& \cite{wl}\\
\hline
$B_c\to \chi_{c0} e\nu$ & 1.27 & 2.52 & 1.55 & 1.69 & $2.60\pm0.73$\\
$B_c\to \chi_{c0} \tau\nu$ & 0.11 & 0.26 & 0.19 & 0.25 & $0.70\pm0.23$\\
$B_c\to \chi_{c1} e\nu$ & 1.18 & 1.40 & 0.94 & 2.21 & $2.09\pm0.60$\\
$B_c\to \chi_{c1} \tau\nu$ & 0.13 & 0.17 & 0.10 & 0.35 & $0.21\pm0.06$\\
$B_c\to \chi_{c2} e\nu$ & 2.27 & 2.92 & 1.89 & 2.73 \\
$B_c\to \chi_{c2} \tau\nu$ & 0.13 & 0.20 & 0.13 & 0.42 \\
$B_c\to h_{c} e\nu$ & 1.38 & 4.42 & 2.40 & 2.51 & $2.03\pm0.57$& $4.2\pm2.1$\\
$B_c\to h_{c} \tau\nu$ & 0.11 & 0.38 & 0.21 & 0.36 & $0.20\pm0.05$&
$0.53\pm0.26$\\
$B_c\to \chi_{c0}' e\nu$ & 0.19 & & & & & $10\pm 6$\\
$B_c\to \chi_{c0}' \tau\nu$ & 0.0089 & & & & & $0.39\pm 0.20$\\
$B_c\to \chi_{c1}' e\nu$ & 0.12 & & & & & $8.6\pm 4.8$\\
$B_c\to \chi_{c1}' \tau\nu$& 0.0056 & & & & & $0.31\pm 0.18$\\
$B_c\to \chi_{c2}' e\nu$& 0.048 & & & & & \\
$B_c\to \chi_{c2}' \tau\nu$ &0.0019 \\
$B_c\to h_{c}' e\nu$ &0.031 & & & & & $0.76\pm 0.33$\\ 
$B_c\to h_{c}' \tau\nu$ &0.0016 & & & & & $0.028\pm 0.014$\\
\end{tabular}
\end{ruledtabular}
\end{table}

\begin{table}
\caption{Predictions for the  rates of the semileptonic $B_c$ decays
  to the $P$-wave $D$ mesons (in
  $10^{-15}$ GeV). }
\label{brbcd}
\begin{ruledtabular}
\begin{tabular}{cccc}
 Decay& $\Gamma$ &Decay& $\Gamma$\\
\hline
$B_c\to D_{0} e\nu$ &0.016 & 
$B_c\to D_{0} \tau\nu$ & 0.0067\\
$B_c\to D_{1} e\nu$ & 0.016 &
$B_c\to D_{1} \tau\nu$ & 0.0056\\
$B_c\to D_{1}' e\nu$ & 0.027 &
$B_c\to D_{1}' \tau\nu$ & 0.016 \\
$B_c\to D_{2} e\nu$ & 0.052 &
$B_c\to D_{2} \tau\nu$ &0.019 
\end{tabular}
\end{ruledtabular}
\end{table}

\begin{figure}
  \centering
 \includegraphics[width=8cm]{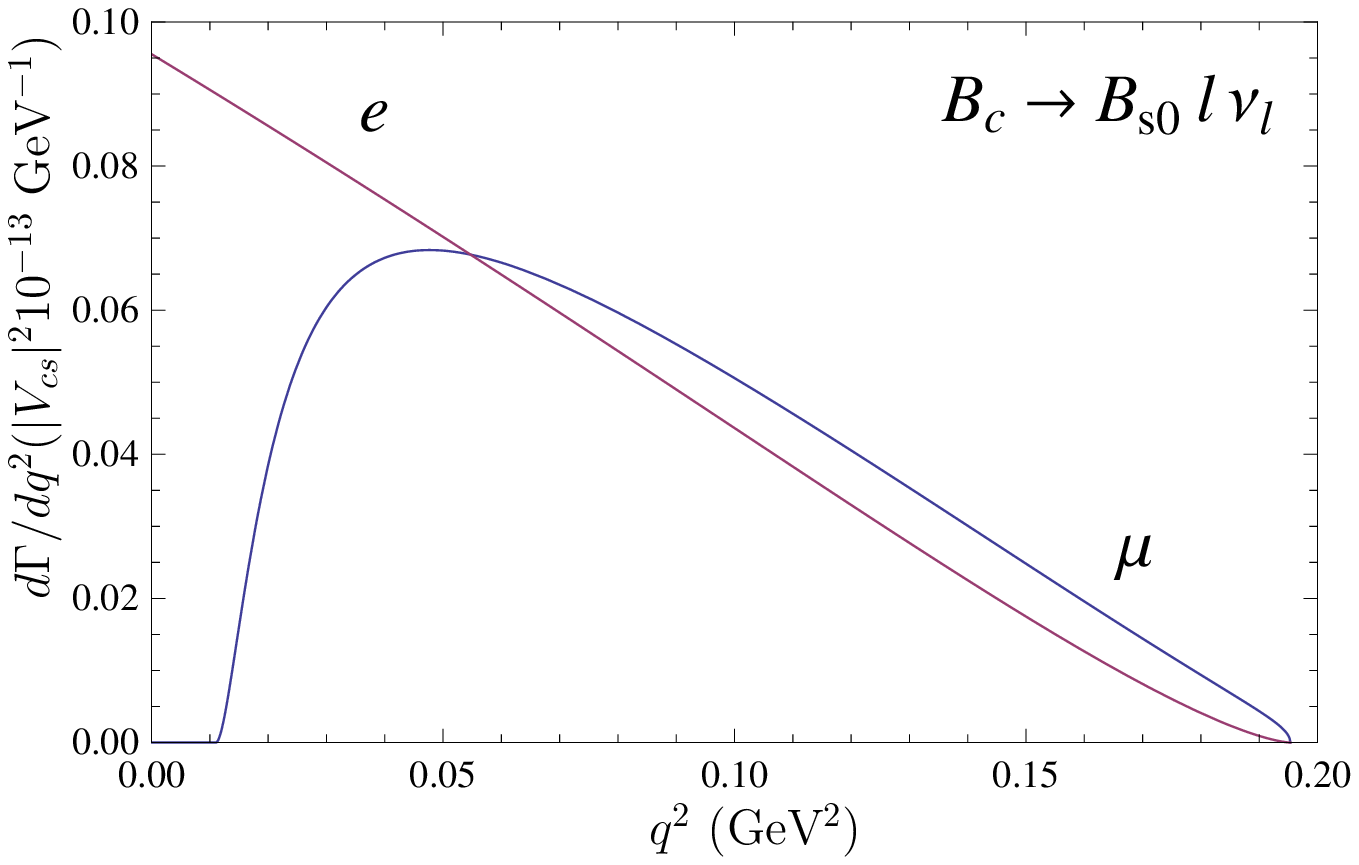}\ \
 \  \includegraphics[width=8cm]{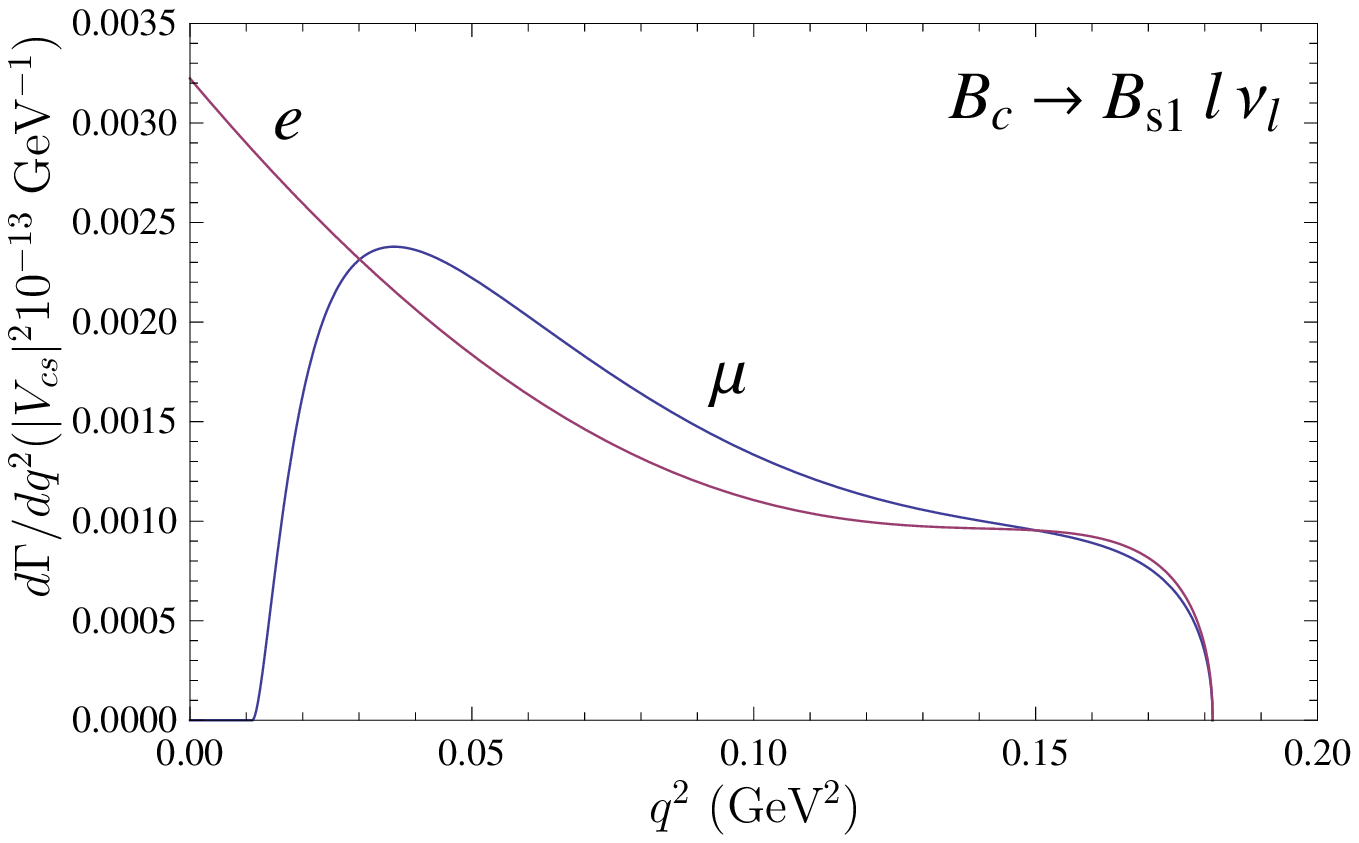}

\vspace*{0.5cm}
 \includegraphics[width=8cm]{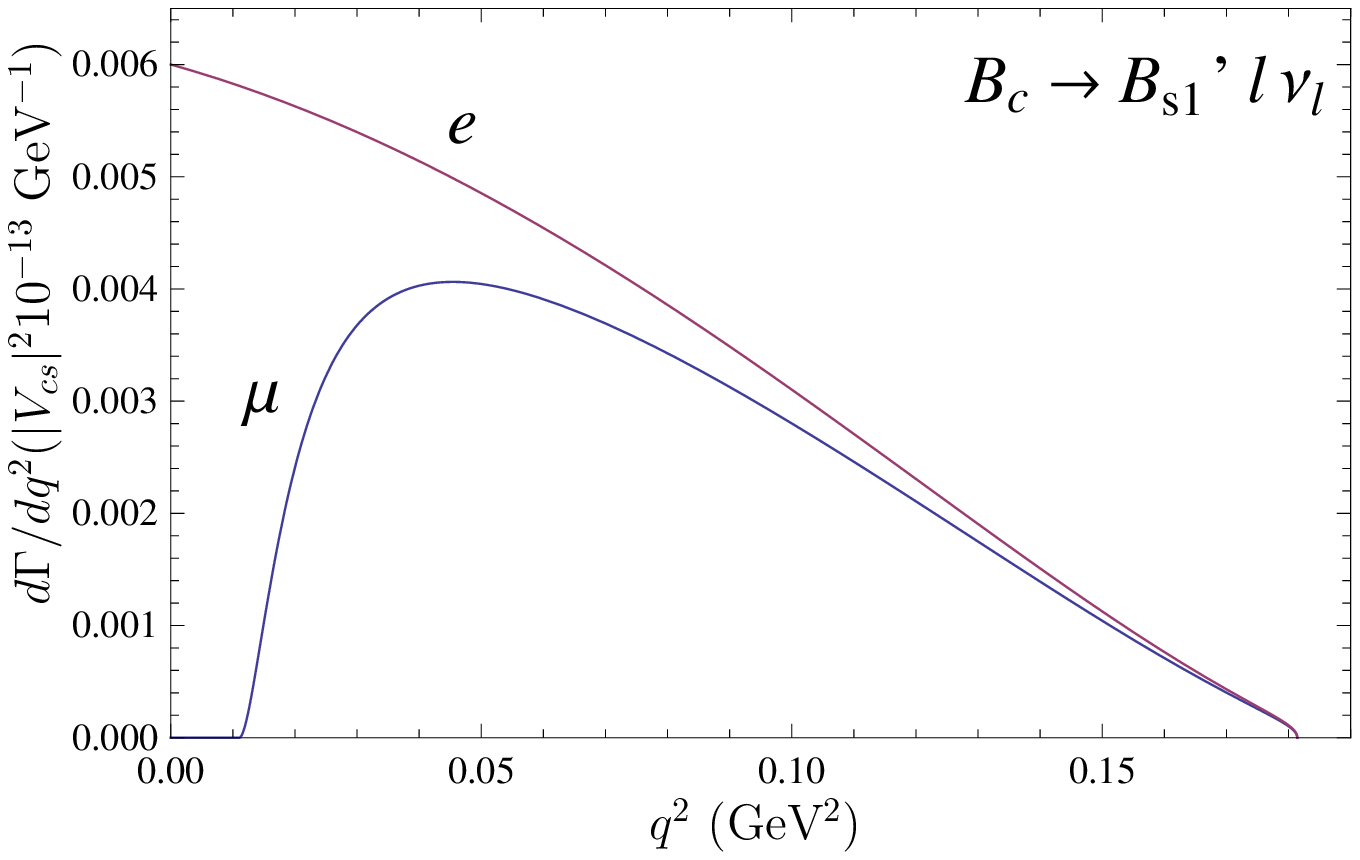}\ \ \  \includegraphics[width=8cm]{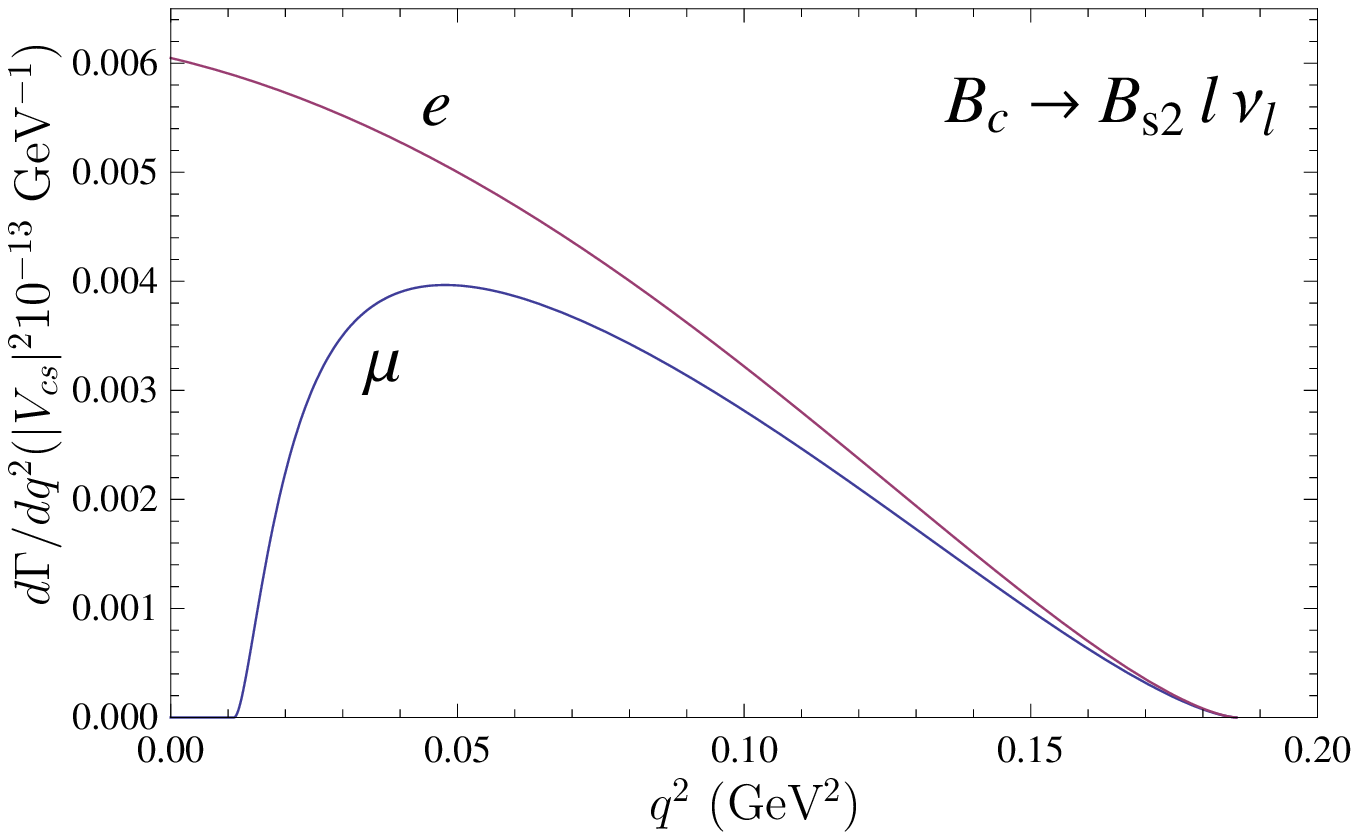}

  \caption{Predictions for the differential decay rates  of the $B_c$ 
    semileptonic decays to the $P$-wave $B_s$ meson states. }
  \label{fig:brbcbs}
\end{figure}

In Fig.~\ref{fig:brbcbs} we plot predicted differential semileptonic
decay rates of the $B_c$ to $P$-wave $B_s$ meson states,
governed by the $c\to s$ weak transitions. The allowed kinimatical
range for these transitions is rather narrow. Therefore semileptonic
decays involving the $\tau$ lepton are forbidden. From these plots we see
that even the account of the rather small muon mass
significantly modifies the
differential decay rates. The corresponding plots for $B_c\to B_Jl\nu$
decays have similar shape and  are not shown here. The predicted
values for the rates of the semileptonic $B_c$ decays to the $P$-wave $B_s$
and $B$ mesons are given in Tables~\ref{brbcbs} and \ref{brbcb}. Note
that, notwithstanding the fact that these decay rates have significantly
larger values of the CKM matrix elements than the rates of
$B_c$ decays to charmonium,  they have the same order of
magnitude. This is the result of the above mentioned strong phase
space suppression of the $B_c\to B_{sJ}l\nu$ decays. 

\begin{table}
\caption{Predictions for the  rates of the semileptonic $B_c$ decays
  to the $P$-wave $B_s$ mesons (in
  $10^{-15}$ GeV). }
\label{brbcbs}
\begin{ruledtabular}
\begin{tabular}{cccc}
 Decay& $\Gamma$ &Decay& $\Gamma$\\
\hline
$B_c\to B_{s0} e\nu$ &0.96 & 
$B_c\to B_{s0} \mu\nu$ & 0.82\\
$B_c\to B_{s1} e\nu$ & 0.029 &
$B_c\to B_{s1} \mu\nu$ & 0.026\\
$B_c\to B_{s1}' e\nu$ & 0.065 &
$B_c\to B_{s1}' \mu\nu$ & 0.044 \\
$B_c\to B_{s2} e\nu$ & 0.066 &
$B_c\to B_{s2} \mu\nu$ &0.031 
\end{tabular}
\end{ruledtabular}
\end{table}

\begin{table}
\caption{Predictions for the rates of the semileptonic $B_c$ decays
  to the $P$-wave $B$ mesons (in
  $10^{-15}$ GeV). }
\label{brbcb}
\begin{ruledtabular}
\begin{tabular}{cccc}
 Decay& $\Gamma$ &Decay& $\Gamma$\\
\hline
$B_c\to B_{0} e\nu$ &0.089 & 
$B_c\to B_{0} \mu\nu$ & 0.082\\
$B_c\to B_{1} e\nu$ & 0.0048 &
$B_c\to B_{1} \mu\nu$ & 0.0043\\
$B_c\to B_{1}' e\nu$ & 0.010 &
$B_c\to B_{1}' \mu\nu$ & 0.0082 \\
$B_c\to B_{2} e\nu$ & 0.012 &
$B_c\to B_{2} \mu\nu$ &0.0067 
\end{tabular}
\end{ruledtabular}
\end{table}

\section{Nonleptonic decays}\label{nl}
In the standard model nonleptonic $B_c$ decays are described by the
effective Hamiltonian, obtained by integrating out the heavy $W$-boson
and top quark. 

(a) For the case of the $b\to c,u$ transitions, one gets
\begin{equation}
\label{heff}
H_{\rm eff}=\frac{G_F}{\sqrt{2}}V_{cb}\left[c_1(\mu)O_1^{cb}+
c_2(\mu)O_2^{cb}\right] 
+\frac{G_F}{\sqrt{2}}V_{ub}\left[c_1(\mu)O_1^{ub}+
c_2(\mu)O_2^{ub}\right] +\dots.
\end{equation}

(b) For the case of the $c\to s,d$ transitions, we have
\begin{equation}
\label{heffc}
H_{\rm eff}=\frac{G_F}{\sqrt{2}}V_{cs}\left[c_1(\mu)O_1^{cs}+
c_2(\mu)O_2^{cs}\right] 
+\frac{G_F}{\sqrt{2}}V_{cd}\left[c_1(\mu)O_1^{cd}+
c_2(\mu)O_2^{cd}\right] +\dots.
\end{equation}
The Wilson coefficients $c_{1,2}(\mu)$ are evaluated
perturbatively at the $W$ scale and then are evolved down to the
renormalization scale $\mu\approx m_b$ by the renormalization-group
equations. The ellipsis denote the penguin operators, the Wilson
coefficients of  which are numerically much smaller than $c_{1,2}$.
The local four-quark operators $O_1$ and $O_2$ are given by
\begin{eqnarray}
\label{o12}
O_1^{qb}&=& [({\tilde d}u)_{V-A}+({\tilde s}c)_{V-A}](\bar q
b)_{V-A}, \cr O_2^{qb}&=& (\bar qu)_{V-A}({\tilde d}b)_{V-A}+(\bar
qc)_{V-A}({\tilde s}b)_{V-A}, \qquad q=(u,c),
\end{eqnarray}
and
\begin{eqnarray}
\label{o12c}
O_1^{cq}&=& ({\tilde d}u)_{V-A}(\bar c
q)_{V-A}, \cr O_2^{cq}&=& (\bar
cu)_{V-A}({\tilde d}q)_{V-A}, \qquad q=(s,d),
\end{eqnarray}
where the rotated antiquark fields are
\begin{equation} \label{ds}
\tilde d=V_{ud}\bar d+V_{us}\bar s,
 \qquad \tilde s=V_{cd}\bar d+V_{cs}\bar s,
\end{equation}
and for
the hadronic current the following notation is used
$$(\bar qq')_{V-A}=\bar q\gamma_\mu(1-\gamma_5)q' \equiv J^W_\mu.$$

The factorization approach, which is extensively used for the calculation
of two-body nonleptonic decays, such as $B_c\to FM$, assumes that the
nonleptonic decay amplitude reduces to the product of a meson
transition matrix element
and a decay constant \cite{bsw}. This assumption in general cannot be
exact.  However, it is expected that factorization can hold 
for energetic decays, where the final $F$ meson is heavy and the $M$
meson is light \cite{dg}. A justification of this assumption is
usually based on the issue of color 
transparency \cite{jb}. In these decays the final hadrons
are produced in the form of almost point-like color-singlet objects with a
large relative momentum. And thus the hadronization of the decay
products occurs  after they are too far separated for strongly interacting
with each other. That provides the possibility to avoid the final
state interaction. A more general treatment of factorization is given in
Refs.~\cite{bbns,bs}.  

\begin{figure}
  \centering 
  \includegraphics[width=7.5cm]{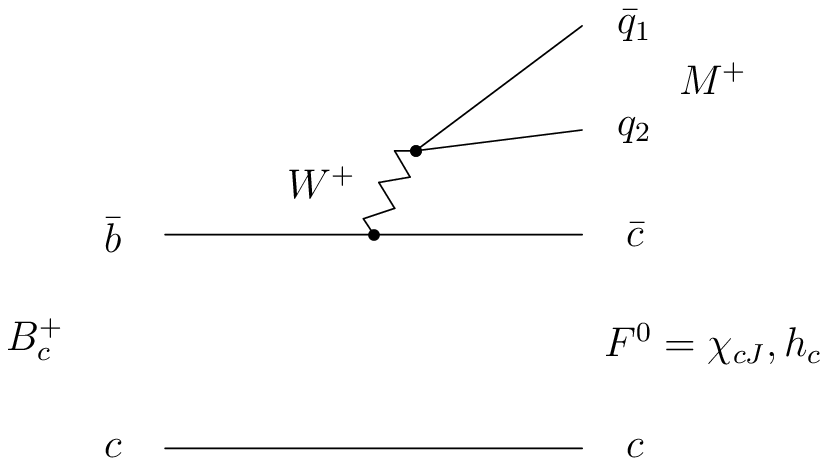}\ \ \ \
  \ \includegraphics[width=7.5cm]{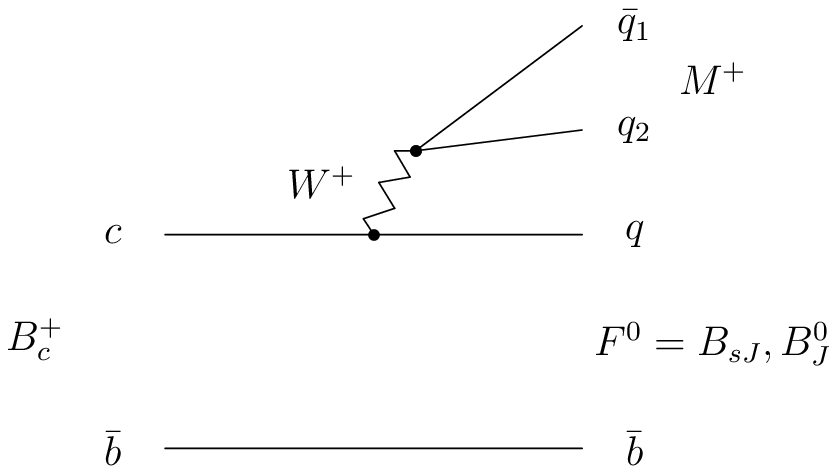}

(a) \hspace*{7cm} (b)\hspace*{1cm}
\caption{Quark diagram for the nonleptonic $B_c^+\to F^0M^+$ decay. }
\label{d3}
\end{figure}

Here we first analyze  the $B_c^+$ nonleptonic decays to
the $P$-wave charmonium and the light $\pi^+$, $\rho^+$ or 
$K^{(*)+}$ mesons, governed by
the weak $b\to c,u$ transitions. The corresponding diagram is shown in
Fig.~\ref{d3}(a), where $q_1=d$,  $s$ and $q_2=u$. Then the decay
amplitude can be approximated by the product of one-particle matrix
elements 
\begin{equation}
\label{factor} \langle F^0M^+|H_{\rm eff}|B_c^+\rangle= \frac{G_F}{\sqrt{2}}
V_{cb}V_{q_1q_2} a_1\langle F|(\bar
bc)_{V-A}|B_c\rangle\langle M|(\bar q_1q_2)_{V-A}|0\rangle ,
\end{equation}
where
\begin{equation}
\label{amu} a_1=c_1(\mu)+\frac{1}{N_c}c_2(\mu)
\end{equation}
and $N_c$ is the number of colors. 

Next we consider  nonleptonic decays of the $B_c$ meson to the $P$-wave
$B_s$ or $B$ mesons and the final light $M^+$ meson, governed by
the weak $c\to s,d$ transitions. Only the 
pion is kinematically allowed. The corresponding diagram is shown in
Fig.~\ref{d3}(b). Then in the
factorization approximation the decay
amplitude can be expressed through the product of one-particle matrix
elements 
\begin{equation}
\label{factor1} 
\langle F^0M^+|H_{\rm eff}|B_c^+\rangle= \frac{G_F}{\sqrt{2}}
V_{cq}V_{q_1q_2} a_1\langle F|(\bar
cq)_{V-A}|B_c\rangle\langle M|(\bar q_1q_2)_{V-A}|0\rangle .
\end{equation}

The matrix element of the weak current $J^W_\mu$ between vacuum and a final
pseudoscalar ($P$) or vector ($V$) meson is parametrized by the decay
constants $f_{P,V}$
\begin{equation}
\langle P|\bar q_1 \gamma^\mu\gamma_5 q_2|0\rangle=if_Pp^\mu_P, \qquad
\langle V|\bar q_1\gamma_\mu q_2|0\rangle=\epsilon_\mu M_Vf_V.
\end{equation}
The pseudoscalar $f_P$ and vector $f_V$ decay constants were
calculated within our model in Ref.~\cite{fpconst}. It was shown that
the complete account of relativistic effects is necessary
to get agreement with experiment for decay constants especially of
light mesons. 
We use the following values of the decay constants: $f_\pi=0.131$~GeV,
$f_\rho=0.208$~GeV, $f_K=0.160$~GeV and $f_{K^*}=0.214$~GeV. The relevant CKM
matrix elements are $|V_{ud}|=0.975$,  $|V_{us}|=0.222$. 

The matrix elements of the weak current between the $B_c$ meson and
the final heavy meson  $F$ entering the factorized nonleptonic decay
amplitude (\ref{factor}) are parametrized by the set of decay form
factors defined in Eqs.~(\ref{eq:f+})-(\ref{eq:tV}). Using the
form factors obtained in Sec.~\ref{ffr}, we get predictions for
the nonleptonic $B_c^+\to \chi_J(h_c)^0M^+$  decay rates
and give them in Table~\ref{nldr} in 
comparison with other calculations \cite{iks,hnv,ccwz,kps}, which are
available for the decays to the $1P$ charmonium states
only. Predictions for the energetic nonleptonic decays to the $2P$
charmonium states are made for the first time and their measurement
could be important for the identification of these states. Our results
for the nonleptonic $B_c^+\to B_{sJ}M^+$ and $B_c^+\to B_{J}M^+$
decay rates are presented in Table~\ref{nldrb}. Note that in the latter case only decays
involving pions are kinematically allowed.

\begin{table}
\caption{The  rates of the nonleptonic $B_c$ decays to the $P$-wave charmonium
  and light mesons (in $10^{-15}$ GeV). }
\label{nldr}
\begin{ruledtabular}
\begin{tabular}{cccccc}
Decay& our&  \cite{iks} & \cite{hnv} & \cite{ccwz}& \cite{kps}\\
\hline
$B_c^+\to\chi_{c0}\pi^+$ & 0.23$a_1^2$ & 0.622$a_1^2$ & 0.28$a_1^2$ &
0.317$a_1^2$& 11$a_1^2$ \\
$B_c^+\to\chi_{c0}\rho^+$ & 0.64$a_1^2$ & 1.47$a_1^2$ & 0.73$a_1^2$ &
0.806$a_1^2$& 37$a_1^2$\\
$B_c^+\to\chi_{c0} K^+$ & 0.018$a_1^2$ & 0.0472$a_1^2$ & 0.022$a_1^2$ &
0.00235$a_1^2$\\
$B_c^+\to\chi_{c0} K^{*+}$ & 0.045$a_1^2$ & 0.0787$a_1^2$ & 0.041$a_1^2$ &
0.00443$a_1^2$\\
$B_c^+\to\chi_{c1}\pi^+$ & 0.22$a_1^2$ & 0.0768$a_1^2$ & 0.0015$a_1^2$ &
0.0815$a_1^2$&0.10$a_1^2$ \\
$B_c^+\to\chi_{c1}\rho^+$ & 0.16$a_1^2$ & 0.326$a_1^2$ & 0.11$a_1^2$ &
0.331$a_1^2$ &5.2$a_1^2$\\
$B_c^+\to\chi_{c1} K^+$ & 0.016$a_1^2$ & 0.0057$a_1^2$ & 0.00012$a_1^2$ &
0.0058$a_1^2$\\
$B_c^+\to\chi_{c1} K^{*+}$ & 0.010$a_1^2$ & 0.0201$a_1^2$ & 0.0080$a_1^2$ &
0.00205$a_1^2$\\
$B_c^+\to\chi_{c2}\pi^+$ & 0.41$a_1^2$ & 0.518$a_1^2$ & 0.24$a_1^2$ &
0.277$a_1^2$ &8.9$a_1^2$\\
$B_c^+\to\chi_{c2}\rho^+$ & 1.18$a_1^2$ & 1.33$a_1^2$ & 0.71$a_1^2$ &
0.579$a_1^2$&36$a_1^2$\\
$B_c^+\to\chi_{c2} K^+$ & 0.031$a_1^2$ & 0.0384$a_1^2$ & 0.018$a_1^2$ &
0.00199$a_1^2$\\
$B_c^+\to\chi_{c2} K^{*+}$ & 0.082$a_1^2$ & 0.0732$a_1^2$ & 0.041$a_1^2$ &
0.00348$a_1^2$\\
$B_c^+\to h_{c}\pi^+$ & 0.51$a_1^2$ & 1.24$a_1^2$ & 0.58$a_1^2$ &
0.569$a_1^2$&18$a_1^2$ \\
$B_c^+\to h_{c}\rho^+$ & 1.11$a_1^2$ & 2.78$a_1^2$ & 1.41$a_1^2$ &
1.40$a_1^2$&60$a_1^2$\\
$B_c^+\to h_{c} K^+$ & 0.039$a_1^2$ & 0.0939$a_1^2$ & 0.045$a_1^2$ &
0.0043$a_1^2$\\
$B_c^+\to h_{c} K^{*+}$ & 0.077$a_1^2$ & 0.146$a_1^2$ & 0.078$a_1^2$ &
0.0076$a_1^2$\\
$B_c^+\to\chi_{c0}'\pi^+$ & 0.023$a_1^2$ & \\
$B_c^+\to\chi_{c0}'\rho^+$ & 0.080$a_1^2$ & \\
$B_c^+\to\chi_{c0}' K^+$ & 0.0019$a_1^2$ &\\
$B_c^+\to\chi_{c0}' K^{*+}$ & 0.0055$a_1^2$ & \\
$B_c^+\to\chi_{c1}'\pi^+$ & 0.011$a_1^2$ & \\
$B_c^+\to\chi_{c1}'\rho^+$ & 0.016$a_1^2$ & \\
$B_c^+\to\chi_{c1}' K^+$ & 0.0095$a_1^2$ & \\
$B_c^+\to\chi_{c1} K^{*+}$ & 0.0011$a_1^2$ &\\
$B_c^+\to\chi_{c2}'\pi^+$ & 8.5$\times 10^{-7}$$a_1^2$ &\\
$B_c^+\to\chi_{c2}'\rho^+$ & 0.0022$a_1^2$ & \\
$B_c^+\to\chi_{c2}' K^+$ & 8.5$\times 10^{-6}$$a_1^2$ & \\
$B_c^+\to\chi_{c2}' K^{*+}$ & 0.00015$a_1^2$ & \\
$B_c^+\to h_{c}'\pi^+$ & 1.0$\times 10^{-5}$$a_1^2$ &  \\
$B_c^+\to h_{c}'\rho^+$ & 0.0051$a_1^2$ &\\
$B_c^+\to h_{c}' K^+$ & 3.4$\times10^{-6}$$a_1^2$ &\\
$B_c^+\to h_{c} K^{*+}$ & 0.00035$a_1^2$ &\\
\end{tabular}
\end{ruledtabular}
\end{table}

\begin{table}
\caption{The  rates of the nonleptonic $B_c$ decays to the $P$-wave $B_s$ or $B$ mesons
  and $\pi$ meson (in $10^{-15}$ GeV). }
\label{nldrb}
\begin{ruledtabular}
\begin{tabular}{cccc}
Decay& $\Gamma$& Decay& $\Gamma$ \\
\hline
$B_c^+\to B_{s0}\pi^+$ & 5.82$a_1^2$ &$B_c^+\to B_{0}^0\pi^+$ &
0.46$a_1^2$\\
$B_c^+\to B_{s1}\pi^+$ & 0.30$a_1^2$ &$B_c^+\to B_{1}^0\pi^+$ &
0.041$a_1^2$\\
$B_c^+\to B_{s1}^{'}\pi^+$ & 0.31$a_1^2$ &$B_c^+\to B_{1}^{'0}\pi^+$ &
0.061$a_1^2$\\
$B_c^+\to B_{s2}\pi^+$ & 0.26$a_1^2$ &$B_c^+\to B_{2}^0\pi^+$ & 0.047$a_1^2$\\
\end{tabular}
\end{ruledtabular}
\end{table}

\section{Conclusions}
\label{sec:concl}

We calculated the form factors of the weak $B_c$ decays
to orbitally excited heavy mesons, governed both by the $b\to c,u$ and
$c\to s,d$ transitions, in the framework of the QCD-motivated
relativistic quark model based on the quasipotential approach. The
momentum dependence of the weak decay form factors was reliably
determined in the whole accessible kinematical range. This is
particularly important for $B_c$ decays to the $P$-wave charmonium and
$D$ mesons since they have a rather broad kinematically allowed range
($q_{\rm max}^2\sim 6-15$~GeV$^2$). All essential relativistic effects
were taken into account including transformations of the meson wave
functions from the rest to the moving reference frame and
contributions from the intermediate negative-energy states. The
resulting form factors are expressed through the overlap integrals of
the meson wave functions. These wave functions were obtained
previously in the meson mass spectra calculations and are used in the
present numerical evaluations. The influence of mixing effects on the $P$-wave
heavy-light meson wave functions due to the non-diagonal spin-orbit
and tensor terms in the $Q\bar q$ quasipotential was explicitly
considered. The reliable determination of the $q^2$ dependence of the
from factors in the whole kinematical range is an important
achievement, since in many  previous
calculations form factors were determined only at the single point of
either zero ($q^2=q^2_{\rm max}$) or maximum ($q^2=0$) recoil of the final
meson, and then different ad hoc extrapolations were employed.

The obtained weak form factors were used for the calculation of the
semileptonic and nonleptonic $B_c$ decays to corresponding orbitally excited heavy
mesons. For the nonleptonic decays the factorization approximation was
used. The calculated branching fractions are summarized in
Table~\ref{Br}. In this table we give our predictions not only for
$B_c$ decays to the first $1P$-wave charmonium states ($\chi_J,h_c)$,
but also for their radial excitations ($2P$-wave charmonium $\chi'_J,h'_c)$.  
For completeness, we also present there our predictions for the
semileptonic $B_c$ decays to the $3S$ charmonium states
($\psi'',\eta_c''$) which were not given in our previous study
\cite{bcjpsi}. These decays to highly (both radially and orbitally)
excited charmonium are of 
special interest, since their observation could help to reveal the
nature of the newly observed charmonium-like states
above the open charm production threshold.

\begin{table}
\caption{Branching fractions (in \%) of exclusive $B_c$ decays calculated
  for the fixed values of the $B_c$ lifetime $\tau_{B_c}=0.46$~ps and
 $a_1=1.14$ for the $b\to c$ transitions and $a_1=1.20$ for the $c\to s,d$ transitions.}
\label{Br}
\begin{ruledtabular}
\begin{tabular}{cccccc}
Decay& Br & Decay & Br & Decay & Br\\
\hline
$B_c\to \chi_{c0} e\nu$& 0.087 &
$B_c^+\to \chi_{c0} \pi^{+} $& 0.021  &
$B_c^+\to \chi'_{c0} \pi^{+} $& 0.0020 \\ 
$B_c\to \chi_{c0} \tau\nu$& 0.0075 &
$B_c^+\to \chi_{c0}\rho^+ $& 0.058 & 
$B_c^+\to \chi'_{c0}\rho^+ $& 0.0071 \\
$B_c\to \chi_{c1} e\nu$& 0.082 &
$B_c^+\to \chi_{c0}K^+ $& 0.0016  &
$B_c^+\to \chi'_{c0}K^+ $& 0.00017\\
$B_c\to \chi_{c1} \tau\nu$& 0.0092 &
$B_c^+\to \chi_{c0}K^{*+}$& 0.0040 &
$B_c^+\to \chi'_{c0}K^{*+}$& 0.00049 \\ 
$B_c\to \chi_{c2} e\nu$& 0.16 & 
$B_c^+\to \chi_{c1}\pi^+$& 0.020 &
$B_c^+\to \chi'_{c1}\pi^+$& 0.0010\\
$B_c\to \chi_{c2} \tau\nu$& 0.0093 &
$B_c^+\to \chi_{c1}\rho^{+}$& 0.015&
$B_c^+\to \chi_{c1}\rho^{+}$& 0.0014\\
$B_c\to h_c e\nu$&0.096 & 
$B_c^+\to \chi_{c1} K^{+}$& 0.0015& 
$B_c^+\to \chi'_{c1} K^{+}$& 0.000086\\
$B_c\to h_c\tau\nu$&0.0077 &
$B_c^+\to \chi_{c1} K^{*+}$& 0.0010  &
$B_c^+\to \chi'_{c1} K^{*+}$& 0.00010\\

$B_c\to \chi'_{c0} e\nu$& 0.014 &
$B_c^+\to \chi_{c2} \pi^{+} $& 0.038  &
$B_c^+\to \chi'_{c2} \pi^{+} $& 7.7$\times10^{-8}$ \\ 
$B_c\to \chi'_{c0} \tau\nu$& 0.00063 &
$B_c^+\to \chi_{c2}\rho^+ $& 0.11 & 
$B_c^+\to \chi'_{c2}\rho^+ $& 0.00020 \\
$B_c\to \chi'_{c1} e\nu$& 0.0085 &
$B_c^+\to \chi_{c2}K^+ $& 0.0028  &
$B_c^+\to \chi'_{c2}K^+ $& 7.8$\times10^{-7}$\\
$B_c\to \chi'_{c1} \tau\nu$& 0.00039 &
$B_c^+\to \chi_{c2}K^{*+}$& 0.0074 &
$B_c^+\to \chi'_{c2}K^{*+}$& 0.000014 \\ 
$B_c\to \chi'_{c2} e\nu$& 0.0033 & 
$B_c^+\to h_c \pi^+$& 0.046& 
$B_c^+\to h'_c \pi^+$&9.4$\times10^{-7}$\\
$B_c\to \chi'_{c2} \tau\nu$& 0.00013 &
$B_c^+\to h_c \rho^{+}$& 0.10&
$B_c^+\to h'_c \rho^{+}$& 0.00046\\
$B_c\to h'_c e\nu$&0.0021 & 
$B_c^+\to h_c K^{+}$& 0.0035& 
$B_c^+\to h'_c K^{+}$& 3.1$\times10^{-7}$\\
$B_c\to h'_c\tau\nu$&0.00011 &
$B_c^+\to h_c K^{*+}$& 0.0070  &
$B_c^+\to h'_c K^{*+}$& 0.000032\\

$B_c\to D_{0} e\nu$& 0.0011 &
$B_c\to B_{s0} e\nu$& 0.0066 & 
$B_c\to B_{0} e\nu$& 0.0061 \\ 
$B_c\to D_{0} \tau\nu$& 0.00046 &
$B_c\to B_{s0} \mu\nu$& 0.0057 &
$B_c\to B_{0} \mu\nu$& 0.0056 \\
$B_c\to D_{1} e\nu$& 0.0011 &
$B_c\to B_{s1} e\nu$& 0.0020 & 
$B_c\to B_{1} e\nu$& 0.00033 \\ 
$B_c\to D_{1} \tau\nu$& 0.00039 &
$B_c\to B_{s1} \mu\nu$& 0.0018 &
$B_c\to B_{1} \mu\nu$& 0.00030\\
$B_c\to D'_{1} e\nu$& 0.0019 &
$B_c\to B'_{s1} e\nu$& 0.0045 &
$B_c\to B'_{1} e\nu$& 0.00072 \\ 
$B_c\to D'_{1} \tau\nu$& 0.0011 &
$B_c\to B'_{s1} \mu\nu$& 0.0031 &
$B_c\to B'_{1} \mu\nu$& 0.00057\\
$B_c\to D_{2} e\nu$& 0.0036 &
$B_c\to B_{s2} e\nu$& 0.0046 &
$B_c\to B_{2} e\nu$& 0.00084  \\ 
$B_c\to D_{2} \tau\nu$& 0.0013 &
$B_c\to B_{s2} \mu\nu$& 0.0022 &
$B_c\to B_{2} \mu\nu$& 0.00047\\

$B_c\to\eta_c''e\nu$&0.00055&
$B_c^+\to B_{s0}\pi^+$& 0.55&
$B_c^+\to B_{0}\pi^+$& 0.043\\
$B_c\to\eta_c''\tau\nu$&5.0$\times10^{-7}$&
$B_c^+\to B_{s1}\pi^+$& 0.028&
$B_c^+\to B_{1}\pi^+$& 0.0039\\
$B_c\to\psi''e\nu$&0.00057&$B_c^+\to B'_{s1}\pi^+$& 0.029&
$B_c^+\to B'_{0}\pi^+$& 0.0058\\
$B_c\to\psi''\tau\nu$&3.6$\times10^{-6}$&$B_c^+\to B_{s2}\pi^+$& 0.024&
$B_c^+\to B_{2}\pi^+$& 0.0044\\
\end{tabular}
\end{ruledtabular}
\end{table}

Summing the corresponding branching fractions in Table~\ref{Br} we
find that the semileptonic\footnote{We also take into account
  semileptonic decays involving the muon, which rates are almost equal to
  the ones with the electron.}  and the considered energetic nonleptonic
decays to the $1P$
charmonium states contribute about 0.88\% and 0.44\% of the total rate,
respectively. The corresponding decays to the $2P$ charmonium states
are significantly suppressed (by an order of magnitude) mainly due
to the presence of the node in the $2P$ wave function and give about
0.057\% and 0.013\% of the total rate. The same pattern was previously observed in $B_c$
decays to the $S$-wave charmonia \cite{bcjpsi}, where the  rates of decays to
the $2S$ states were also suppressed by an order of magnitude compared to
decays to the $1S$ states. For decays to higher charmonium excitations such
suppression should be even more pronounced. The CKM suppressed
semileptonic decays to the $D_J$ mesons contribute about 0.019\%.
Thus the total contribution of the considered $B_c$ decays, governed
by the  $b\to c,u$ weak transitions, is about 1.41\%. 

The $B_c$ semileptonic
decays to orbitally excited $B_{sJ}$ and $B_J$ mesons, governed by
the $c\to s,d$ weak transitions, turn out to have smaller branching
fractions than $B_c$ decays to orbitally excited charmonium,
notwithstanding the significantly larger values of the CKM matrix elements,
due to the substantial phase space suppression. Such decays involving the
$\tau$ are kinematically forbidden, while the muon mass starts to play
an important role, reducing branching fractions by more than 10\%. In
total, such semileptonic decays give about 0.045\% of the $B_c$ decay
rate. On the other hand, there is no kinematical suppression in
the corresponding nonleptonic decays and they contribute about
0.69\%. Thus, the total contribution of the considered $B_c$ decays, governed
by the  $c\to s,d$ weak transitions, is about 0.73\%.  

The semileptonic and nonleptonic $B_c$ decays to
excited heavy mesons
can be investigated at Tevatron and LHC, especially in the
LHCb experiment, where the $B_c$ mesons are expected to be copiously
produced.  

\acknowledgements
The authors are grateful to M. Ivanov, V. Matveev,
M. M\"uller-Preussker and 
V. Savrin for support and discussions.  
This work was supported in part by  the Deutsche
Forschungsgemeinschaft under contract Eb 139/4-1 and the Russian
Foundation for Basic Research (RFBR) grants
No.08-02-00582 and No.10-02-91339.

\appendix*
\section{Form factors of weak $B_c$ decays to orbitally excited
  heavy mesons}

(a) $B_c\to S(^3P_0)$ transition ($S=\chi_{c0},D_0^*$)
\nopagebreak
\begin{eqnarray}
  \label{eq:fpl}
  f_\pm^{(1)}(q^2)&=&\sqrt{\frac{E_S}{M_{B_c}}}\int \frac{{\rm
  d}^3p}{(2\pi)^3} \bar\psi_S\left({\bf p}+\frac{2m_c}{E_S+M_S}{\bf
  \Delta} \right)\sqrt{\frac{\epsilon_q(p+\Delta)+
  m_q}{2\epsilon_q(p+\Delta)}} \sqrt{\frac{\epsilon_b(p)+
  m_b}{2\epsilon_b(p)}}\cr
&&\times\Biggl\{\frac{({\bf
p\Delta})}{p{\bf \Delta}^2}\left[\frac{{\bf
\Delta}^2}{\epsilon_q(p+\Delta)+m_q}\pm(M_{B_c}\mp E_S)\left(1+
\frac{{\bf p}^2}{[\epsilon_q(p+\Delta)+m_q][\epsilon_b(p)+m_b]}\right)\right] \cr
&&+\frac23\frac{p}{E_S+M_S}\left(\frac1{\epsilon_q(p+\Delta)+m_q}-
\frac1{\epsilon_c(p)+m_c}\right)\Biggl[\frac{{\bf
\Delta}^2}{\epsilon_q(p+\Delta)+m_q}\pm(M_{B_c}\mp E_S)\cr
&&\times\left(1-
\frac{{\bf
    p}^2}{[\epsilon_q(p+\Delta)+m_q][\epsilon_b(p)+m_b]}\right)\Biggr]
+p\Biggl(\frac1{\epsilon_q(p+\Delta)+m_q}+\frac1{\epsilon_b(p)+m_b}\cr
&&\pm \frac{M_{B_c}\mp E_S}{[\epsilon_q(p+\Delta)+m_q][\epsilon_b(p)+m_b]}\Biggr)\Biggr\}\psi_{B_c}({\bf p}),  
\end{eqnarray}

\begin{eqnarray}
  \label{eq:fpls}
  f_\pm^{S(2)}(q^2)&=&\sqrt{\frac{E_S}{M_{B_c}}}\int \frac{{\rm
  d}^3p}{(2\pi)^3} \bar\psi_S\left({\bf p}+\frac{2m_c}{E_S+M_S}{\bf
  \Delta} \right)\sqrt{\frac{\epsilon_q(\Delta)+
  m_q}{2\epsilon_q(\Delta)}} \Biggl\{
-\frac{({\bf
p\Delta})}{p{\bf \Delta}^2}\frac{{\bf
\Delta}^2}{\epsilon_q(\Delta)[\epsilon_q(p+\Delta)]}\cr
&& \times \Biggl(1\mp\frac{M_{B_c}\mp E_S}{\epsilon_q(\Delta)+mq}\Biggr)
\left[M_S-\epsilon_q\left(p+\frac{2m_c}{E_S+M_S}\Delta\right)
-\epsilon_c\left(p+\frac{2m_c}{E_S+M_S}\Delta \right)\right]\cr
&&
  -p\Biggl(\frac1{4m_b^2}+\frac1{2\epsilon_q(\Delta)[\epsilon_q(\Delta)+m_q]}\Biggr)
\Biggl[M_{B_c}+M_S-\epsilon_b(p)-\epsilon_c(p)\cr
&&
-\epsilon_q\left(p+\frac{2m_c}{E_S+M_S}\Delta\right)-\epsilon_c\left(p+\frac{2m_c}{E_S+M_S}\Delta \right)\Biggr]
+p\frac{\epsilon_q(\Delta)-m_q}{2m_b\epsilon_q(\Delta)[\epsilon_q(\Delta)+m_q]}
 \cr
&&\times\left[M_S-\epsilon_q\left(p+\frac{2m_c}{E_S+M_S}\Delta\right)
-\epsilon_c\left(p+\frac{2m_c}{E_S+M_S}\Delta \right)\right]
\Biggrb)\Biggr\}\psi_{B_c}({\bf p}),
\end{eqnarray}

\begin{eqnarray}
  \label{eq:fplv}
 f_\pm^{V(2)}(q^2)&=&\sqrt{\frac{E_S}{M_{B_c}}}\int \frac{{\rm
  d}^3p}{(2\pi)^3} \bar\psi_S\left({\bf p}+\frac{2m_c}{E_S+M_S}{\bf
  \Delta} \right)\sqrt{\frac{\epsilon_q(\Delta)+
  m_q}{2\epsilon_q(\Delta)}} \cr
&& \times
\frac{p}{2m_c}\Biggl\{
\frac{1}{\epsilon_q(p+\Delta)+m_q} \Biggl(1\mp\frac{M_{B_c}\mp E_S}{\epsilon_q(\Delta)+m_q}\Biggr)+\frac1{2m_b}\Biggl(1\pm\frac{M_{B_c}\mp E_S}{\epsilon_q(\Delta)+m_q}\Biggr)
\cr
&&
-\frac{\epsilon_q(\Delta)-m_q}{3\epsilon_q(\Delta)[\epsilon_q(\Delta)+m_q]}\Biggl(1\mp\frac{M_{B_c}\mp E_S}{\epsilon_q(\Delta)+m_q}\Biggr)\Biggr\}
\Biggl[M_{B_c}+M_S-\epsilon_b(p)-\epsilon_c(p)\cr
&&
-\epsilon_q\left(p+\frac{2m_c}{E_S+M_S}\Delta\right)-\epsilon_c\left(p+\frac{2m_c}{E_S+M_S}\Delta \right)\Biggr]\psi_{B_c}({\bf p}),
\end{eqnarray}

(b) $B_c\to AV(^3P_1)$ transition ($AV=\chi_{c1},D_1(^3P_1)$)

\begin{eqnarray}
  \label{eq:hv1}
  h_{V_1}^{(1)}(q^2)&=&\frac{2\sqrt{E_{AV}M_{B_c}}}{M_{B_c}+M_{AV}}\int \frac{{\rm
  d}^3p}{(2\pi)^3} \bar\psi_{AV}\left({\bf p}+\frac{2m_c}{E_{AV}+M_{AV}}{\bf
  \Delta} \right)\sqrt{\frac{\epsilon_q(p+\Delta)+
  m_q}{2\epsilon_q(p+\Delta)}} \sqrt{\frac{\epsilon_b(p)+
  m_b}{2\epsilon_b(p)}}\cr
&&\times\Biggl\{\frac{({\bf
p\Delta})}{p}\frac{1}{\epsilon_q(p+\Delta)+m_q} 
+p\Biggl[\frac{E_{AV}- M_{AV}}{\epsilon_q(p+\Delta)+m_q} \left(\frac1{\epsilon_q(p+\Delta)+m_q}-\frac1{\epsilon_c(p)+m_c}\right)\cr
&&+ \frac23\left(\frac1{\epsilon_q(p+\Delta)+m_q}-\frac1{\epsilon_b(p)+m_b}\right)\Biggr]\Biggr\}\psi_{B_c}({\bf p}),  
\end{eqnarray}

\begin{eqnarray}
 \label{eq:hv1s}
  h_{V_1}^{S(2)}(q^2)&=&\frac{2\sqrt{E_{AV}M_{B_c}}}{M_{B_c}+M_{AV}}\int \frac{{\rm
  d}^3p}{(2\pi)^3} \bar\psi_{AV}\left({\bf p}+\frac{2m_c}{E_{AV}+M_{AV}}{\bf
  \Delta} \right)\sqrt{\frac{\epsilon_q(\Delta)+
  m_q}{2\epsilon_q(\Delta)}} \cr
&&\times\Biggl\{-\frac{({\bf
p\Delta})}{p}\frac{1}{\epsilon_q(\Delta)[\epsilon_q(\Delta)+m_q]}
\left[M_{B_c}-\epsilon_b(p)-\epsilon_c(p)\right]\cr
&&
-\frac{p}3\Biggl[\Biggl(\frac1{\epsilon_q(\Delta)[\epsilon_q(\Delta)+m_q]}-\frac1{2m_b^2}\Biggr)\Bigl[M_{B_c}+M_{AV}-\epsilon_b(p)-\epsilon_c(p)\cr
&&
-\epsilon_q\left(p+\frac{2m_c}{E_{AV}+M_{AV}}\Delta\right)-\epsilon_c\left(p+\frac{2m_c}{E_{AV}+M_{AV}}\Delta \right)\Bigr]
+\frac{\epsilon_q(\Delta)-m_q}{2m_b\epsilon_q(\Delta)[\epsilon_q(\Delta)+m_q]}
 \cr
&&\times\left[M_{AV}-\epsilon_q\left(p+\frac{2m_c}{E_{AV}+M_{AV}}\Delta\right)
-\epsilon_c\left(p+\frac{2m_c}{E_{AV}+M_{AV}}\Delta \right)\right]
\Biggr]\Biggr\}\psi_{B_c}({\bf p}),  
\end{eqnarray}

\begin{eqnarray}
 \label{eq:hv1v}
  h_{V_1}^{V(2)}(q^2)&=&\frac{2\sqrt{E_{AV}M_{B_c}}}{M_{B_c}+M_{AV}}\int \frac{{\rm
  d}^3p}{(2\pi)^3} \bar\psi_{AV}\left({\bf p}+\frac{2m_c}{E_{AV}+M_{AV}}{\bf
  \Delta} \right)\sqrt{\frac{\epsilon_q(\Delta)+
  m_q}{2\epsilon_q(\Delta)}} \cr
&&\times\frac{p}{3m_c}\Biggl\{\frac{1}{\epsilon_q(\Delta)+m_q}\Bigl[M_{AV}
-\epsilon_q\left(p+\frac{2m_c}{E_{AV}+M_{AV}}\Delta\right)-\epsilon_c\left(p+\frac{2m_c}{E_{AV}+M_{AV}}\Delta \right)\Bigr]\cr
&&
-\frac{m_c}{\epsilon_q(\Delta)[\epsilon_q(\Delta)+m_q]}\left[M_{B_c}-\epsilon_b(p)-\epsilon_c(p)\right]
-\frac{1}{2m_b}\Bigl[M_{B_c}+M_{AV}-\epsilon_b(p)-\epsilon_c(p)\cr
&&
-\epsilon_q\left(p+\frac{2m_c}{E_{AV}+M_{AV}}\Delta\right)-\epsilon_c\left(p+\frac{2m_c}{E_{AV}+M_{AV}}\Delta \right)\Bigr]\Biggr\}\psi_{B_c}({\bf p}),  
\end{eqnarray}

\begin{eqnarray}
  \label{eq:hv2}
  h_{V_2}^{(1)}(q^2)&=&2E_{AV}\sqrt{\frac{E_{AV}}{M_{B_c}}}\int \frac{{\rm
  d}^3p}{(2\pi)^3} \bar\psi_{AV}\left({\bf p}+\frac{2m_c}{E_{AV}+M_{AV}}{\bf
  \Delta} \right)\sqrt{\frac{\epsilon_q(p+\Delta)+
  m_q}{2\epsilon_q(p+\Delta)}} \sqrt{\frac{\epsilon_b(p)+
  m_b}{2\epsilon_b(p)}}\cr
&&\times\Biggl\{\frac{({\bf
p\Delta})}{p{\bf
  \Delta}^2}\frac{E_{AV}}{\epsilon_q(p+\Delta)+m_q}\left[\frac{M_{AV}^2}{E_{AV}^2}
-\frac23\frac{{\bf p}^2}{E_{AV}+
  M_{AV}}\left(\frac1{\epsilon_q(p+\Delta)+m_q}-\frac1{\epsilon_c(p)+m_c}\right)
\right]\cr
&&-\frac23\frac{p}{E_{AV}+
  M_{AV}}\left(\frac1{\epsilon_q(p+\Delta)+m_q}-\frac1{\epsilon_c(p)+m_c}\right)
\Biggl(1-\frac{E_{AV}}{2[\epsilon_q(p+\Delta)+m_q]}\cr
&& +\frac{{\bf p}^2}{[\epsilon_q(p+\Delta)+m_q][\epsilon_b(p)+m_b]}
 +\frac32\frac{{\bf
     \Delta}^2}{E_{AV}[\epsilon_q(p+\Delta)+m_q]}\Biggr)\cr
&&
+\frac23
p\Biggl[\frac{1}{[\epsilon_q(p+\Delta)+m_q][\epsilon_b(p)+m_b]} 
+\cr
&&  
\frac1{E_{AV}}\left(\frac1{\epsilon_q(p+\Delta)+m_q}-\frac1{\epsilon_b(p)+m_b}\right)\Biggr]\Biggr\}\psi_{B_c}({\bf
  p}),
\end{eqnarray}

\begin{eqnarray}
 \label{eq:hv2s}
  h_{V_2}^{S(2)}(q^2)&=&2E_{AV}\sqrt{\frac{E_{AV}}{M_{B_c}}}\int \frac{{\rm
  d}^3p}{(2\pi)^3} \bar\psi_{AV}\left({\bf p}+\frac{2m_c}{E_{AV}+M_{AV}}{\bf
  \Delta} \right)\sqrt{\frac{\epsilon_q(\Delta)+
  m_q}{2\epsilon_q(\Delta)}} \cr
&&\times\Biggl\{-\left(\frac{({\bf
p\Delta})}{p{\bf\Delta}^2}\frac{M_{AV}^2}{E_{AV}}+\frac{2p}{3[\epsilon_q(\Delta)+m_q]}\right)\frac1{\epsilon_q(\Delta)[\epsilon_q(\Delta)+m_q]}
\left[M_{B_c}-\epsilon_b(p)-\epsilon_c(p)\right]\cr
&&
+\frac{p}3\left(\frac1{E_{AV}}-\frac1{\epsilon_q(\Delta)+m_q}\right)\Biggl[\Biggl(\frac1{\epsilon_q(\Delta)[\epsilon_q(\Delta)+m_q]}-\frac1{2m_b^2}\Biggr)\Bigl[M_{B_c}+M_{AV}-\epsilon_b(p)\cr
&&-\epsilon_c(p)
-\epsilon_q\left(p+\frac{2m_c}{E_{AV}+M_{AV}}\Delta\right)-\epsilon_c\left(p+\frac{2m_c}{E_{AV}+M_{AV}}\Delta \right)\Bigr]
+\frac{1}{m_b\epsilon_q(\Delta)}
 \cr
&&\times\left[M_{AV}-\epsilon_q\left(p+\frac{2m_c}{E_{AV}+M_{AV}}\Delta\right)
-\epsilon_c\left(p+\frac{2m_c}{E_{AV}+M_{AV}}\Delta \right)\right]
\Biggr]\Biggr\}\psi_{B_c}({\bf p}),  
\end{eqnarray}

\begin{eqnarray}
 \label{eq:hv2v}
  h_{V_2}^{V(2)}(q^2)&=&2E_{AV}\sqrt{\frac{E_{AV}}{M_{B_c}}}\int \frac{{\rm
  d}^3p}{(2\pi)^3} \bar\psi_{AV}\left({\bf p}+\frac{2m_c}{E_{AV}+M_{AV}}{\bf
  \Delta} \right)\sqrt{\frac{\epsilon_q(\Delta)+
  m_q}{2\epsilon_q(\Delta)}} \cr
&&\times\frac{p}{3m_c}\Biggl\{\Biggl[\frac{1}{[\epsilon_q(\Delta)+m_q]^2}\left(1-\frac{\epsilon_q(\Delta)+m_q}{E_{AV}}-\frac{E_{AV}}{2\epsilon_q(\Delta)}\right)\cr
&&+\frac1{2m_b}\left(\frac1{\epsilon_q(\Delta)+m_q}+\frac1{E_{AV}}\right)\Biggr]
\Bigl[M_{B_c}+M_{AV}-\epsilon_b(p)\cr
&&-\epsilon_c(p)
-\epsilon_q\left(p+\frac{2m_c}{E_{AV}+M_{AV}}\Delta\right)-\epsilon_c\left(p+\frac{2m_c}{E_{AV}+M_{AV}}\Delta
\right)\Bigr]\cr
&&+\frac{1}{E_{AV}\epsilon_q(\Delta)}\left[M_{B_c}-\epsilon_b(p)-\epsilon_c(p)\right]
\Biggr\}\psi_{B_c}({\bf p}),  
\end{eqnarray}

\begin{eqnarray}
  \label{eq:hv3}
  h_{V_3}^{(1)}(q^2)&=&2E_{AV}\sqrt{E_{AV}M_{B_c}}\int \frac{{\rm
  d}^3p}{(2\pi)^3} \bar\psi_{AV}\left({\bf p}+\frac{2m_c}{E_{AV}+M_{AV}}{\bf
  \Delta} \right)\sqrt{\frac{\epsilon_q(p+\Delta)+
  m_q}{2\epsilon_q(p+\Delta)}} \sqrt{\frac{\epsilon_b(p)+
  m_b}{2\epsilon_b(p)}}\cr
&&\times\Biggl\{-\frac1{\epsilon_q(p+\Delta)+m_q}\Biggl(\frac{({\bf
p\Delta})}{p{\bf \Delta}^2}\Biggl[1-\frac23\frac{{\bf q}^2}{E_{AV}+ M_{AV}} \left(\frac1{\epsilon_q(p+\Delta)+m_q}-\frac1{\epsilon_c(p)+m_c}\right)\Biggr]\cr
&&+ \frac{q}3\frac1{E_{AV}+M_{AV}}\left(\frac1{\epsilon_q(p+\Delta)+m_q}-\frac1{\epsilon_c(p)+m_c}\right)\Biggr)\Biggr\}\psi_{B_c}({\bf p}),  
\end{eqnarray}

\begin{eqnarray}
 \label{eq:hv3s}
  h_{V_3}^{S(2)}(q^2)&=&2E_{AV}\sqrt{E_{AV}M_{B_c}}\int \frac{{\rm
  d}^3p}{(2\pi)^3} \bar\psi_{AV}\left({\bf p}+\frac{2m_c}{E_{AV}+M_{AV}}{\bf
  \Delta} \right)\sqrt{\frac{\epsilon_q(\Delta)+
  m_q}{2\epsilon_q(\Delta)}} \cr
&&\times\frac{({\bf
p\Delta})}{p{\bf\Delta}^2}\frac{1}{\epsilon_q(\Delta)[\epsilon_q(\Delta)+m_q]}
\left[M_{B_c}-\epsilon_b(p)-\epsilon_c(p)\right]\psi_{B_c}({\bf p}),  
\end{eqnarray}

\begin{eqnarray}
 \label{eq:hv3v}
  h_{V_3}^{V(2)}(q^2)&=&2E_{AV}\sqrt{E_{AV}M_{B_c}}\int \frac{{\rm
  d}^3p}{(2\pi)^3} \bar\psi_{AV}\left({\bf p}+\frac{2m_c}{E_{AV}+M_{AV}}{\bf
  \Delta} \right)\sqrt{\frac{\epsilon_q(\Delta)+
  m_q}{2\epsilon_q(\Delta)}} \cr
&&\times\frac{p}{6m_c\epsilon_q(\Delta)[\epsilon_q(\Delta)+m_q]^2}\Bigl[M_{B_c}+M_{AV}-\epsilon_b(p)-\epsilon_c(p)\cr
&&
-\epsilon_q\left(p+\frac{2m_c}{E_{AV}+M_{AV}}\Delta\right)-\epsilon_c\left(p+\frac{2m_c}{E_{AV}+M_{AV}}\Delta \right)\Bigr]\psi_{B_c}({\bf p}),  
\end{eqnarray}

\begin{eqnarray}
  \label{eq:ha}
  h_{A}^{(1)}(q^2)&=&(M_{B_c}+M_{AV})\sqrt{\frac{E_{AV}}{M_{B_c}}}\int \frac{{\rm
  d}^3p}{(2\pi)^3} \bar\psi_{AV}\left({\bf p}+\frac{2m_c}{E_{AV}+M_{AV}}{\bf
  \Delta} \right)\sqrt{\frac{\epsilon_q(p+\Delta)+
  m_q}{2\epsilon_q(p+\Delta)}}\cr
&&\times \sqrt{\frac{\epsilon_b(p)+
  m_b}{2\epsilon_b(p)}}\Biggl\{\frac{({\bf
p\Delta})}{p{\bf\Delta}^2}
+\frac{p}3\Biggl[\frac{1}{E_{AV}+ M_{AV}} \left(\frac1{\epsilon_q(p+\Delta)+m_q}-\frac1{\epsilon_c(p)+m_c}\right)\cr
&&+ \frac2{[\epsilon_q(p+\Delta)+m_q][\epsilon_b(p)+m_b]}\Biggr]\Biggr\}\psi_{B_c}({\bf p}),  
\end{eqnarray}

\begin{eqnarray}
 \label{eq:h_as}
  h_{A}^{S(2)}(q^2)&=&(M_{B_c}+M_{AV})\sqrt{\frac{E_{AV}}{M_{B_c}}}\int \frac{{\rm
  d}^3p}{(2\pi)^3} \bar\psi_{AV}\left({\bf p}+\frac{2m_c}{E_{AV}+M_{AV}}{\bf
  \Delta} \right)\sqrt{\frac{\epsilon_q(\Delta)+
  m_q}{2\epsilon_q(\Delta)}} \cr
&&\times\Biggl\{\left([\epsilon_q(\Delta)-m_q]\frac{({\bf
p\Delta})}{p{\bf\Delta}^2}+\frac{p}{3[\epsilon_q(\Delta)+m_q]}\right)\frac{1}{\epsilon_q(\Delta)[\epsilon_q(\Delta)+m_q]}\cr
&&
\times\left[M_{B_c}-\epsilon_b(p)-\epsilon_c(p)\right]
+\frac{p}{3m_b[\epsilon_q(\Delta)+m_q]}\Biggl(\frac1{4m_b}\Bigl[M_{B_c}+M_{AV}-\epsilon_b(p)-\epsilon_c(p)\cr
&&
-\epsilon_q\left(p+\frac{2m_c}{E_{AV}+M_{AV}}\Delta\right)-\epsilon_c\left(p+\frac{2m_c}{E_{AV}+M_{AV}}\Delta \right)\Bigr]
+\frac{1}{\epsilon_q(\Delta)}
 \cr
&&\times\left[M_{AV}-\epsilon_q\left(p+\frac{2m_c}{E_{AV}+M_{AV}}\Delta\right)
-\epsilon_c\left(p+\frac{2m_c}{E_{AV}+M_{AV}}\Delta \right)\right]
\Biggr)\Biggr\}\psi_{B_c}({\bf p}),  \cr &&
\end{eqnarray}

\begin{eqnarray}
 \label{eq:hav}
  h_{A}^{V(2)}(q^2)&=&(M_{B_c}+M_{AV})\sqrt{\frac{E_{AV}}{M_{B_c}}}\int \frac{{\rm
  d}^3p}{(2\pi)^3} \bar\psi_{AV}\left({\bf p}+\frac{2m_c}{E_{AV}+M_{AV}}{\bf
  \Delta} \right)\sqrt{\frac{\epsilon_q(\Delta)+
  m_q}{2\epsilon_q(\Delta)}} \cr
&&\times\left(-\frac{p}{6m_c[\epsilon_q(\Delta)+m_q]}\right)
\left(\frac1{\epsilon_q(\Delta)}+\frac1{m_b}\right)\Bigl[M_{B_c}+M_{AV}-\epsilon_b(p)-\epsilon_c(p)\cr
&&
-\epsilon_q\left(p+\frac{2m_c}{E_{AV}+M_{AV}}\Delta\right)-\epsilon_c\left(p+\frac{2m_c}{E_{AV}+M_{AV}}\Delta \right)\Bigr]\psi_{B_c}({\bf p}),  
\end{eqnarray}

(c) $B_c\to AV'(^1P_1)$ transition ($AV'=h_{c},D_1(^1P_1)$)
\begin{eqnarray}
  \label{eq:gv1}
  g_{V_1}^{(1)}(q^2)&=&\frac{2\sqrt{E_{AV}M_{B_c}}}{M_{B_c}+M_{AV}}\int \frac{{\rm
  d}^3p}{(2\pi)^3} \bar\psi_{AV}\left({\bf p}+\frac{2m_c}{E_{AV}+M_{AV}}{\bf
  \Delta} \right)\sqrt{\frac{\epsilon_q(p+\Delta)+
  m_q}{2\epsilon_q(p+\Delta)}} \sqrt{\frac{\epsilon_b(p)+
  m_b}{2\epsilon_b(p)}}\cr
&&\times\frac{p}3 \left(\frac1{\epsilon_q(p+\Delta)+m_q}+\frac1{\epsilon_b(p)+m_b}\right)\psi_{B_c}({\bf p}),  
\end{eqnarray}

\begin{eqnarray}
 \label{eq:gv1s}
  g_{V_1}^{S(2)}(q^2)&=&\frac{2\sqrt{E_{AV}M_{B_c}}}{M_{B_c}+M_{AV}}\int \frac{{\rm
  d}^3p}{(2\pi)^3} \bar\psi_{AV}\left({\bf p}+\frac{2m_c}{E_{AV}+M_{AV}}{\bf
  \Delta} \right)\sqrt{\frac{\epsilon_q(\Delta)+
  m_q}{2\epsilon_q(\Delta)}} \cr
&&\times\Biggl\{
-\frac{p}6\Biggl(\frac1{\epsilon_q(\Delta)[\epsilon_q(\Delta)+m_q]}+\frac1{2m_b^2}\Biggr)\Bigl[M_{B_c}+M_{AV}-\epsilon_b(p)-\epsilon_c(p)\cr
&&
-\epsilon_q\left(p+\frac{2m_c}{E_{AV}+M_{AV}}\Delta\right)-\epsilon_c\left(p+\frac{2m_c}{E_{AV}+M_{AV}}\Delta \right)\Bigr]\Biggr\}\psi_{B_c}({\bf p}),  
\end{eqnarray}

\begin{eqnarray}
 \label{eq:gv1v}
  g_{V_1}^{V(2)}(q^2)&=&\frac{2\sqrt{E_{AV}M_{B_c}}}{M_{B_c}+M_{AV}}\int \frac{{\rm
  d}^3p}{(2\pi)^3} \bar\psi_{AV}\left({\bf p}+\frac{2m_c}{E_{AV}+M_{AV}}{\bf
  \Delta} \right)\sqrt{\frac{\epsilon_q(\Delta)+
  m_q}{2\epsilon_q(\Delta)}} \cr
&&\times\frac{p}{6m_c}\left(\frac{1}{\epsilon_q(\Delta)+m_q}
+\frac{1}{2m_b}\right)\Bigl[M_{B_c}+M_{AV}-\epsilon_b(p)-\epsilon_c(p)\cr
&&
-\epsilon_q\left(p+\frac{2m_c}{E_{AV}+M_{AV}}\Delta\right)-\epsilon_c\left(p+\frac{2m_c}{E_{AV}+M_{AV}}\Delta \right)\Bigr]\Biggr\}\psi_{B_c}({\bf p}),  
\end{eqnarray}

\begin{eqnarray}
  \label{eq:gv2}
  g_{V_2}^{(1)}(q^2)&=&2E_{AV}\sqrt{\frac{E_{AV}}{M_{B_c}}}\int \frac{{\rm
  d}^3p}{(2\pi)^3} \bar\psi_{AV}\left({\bf p}+\frac{2m_c}{E_{AV}+M_{AV}}{\bf
  \Delta} \right)\sqrt{\frac{\epsilon_q(p+\Delta)+
  m_q}{2\epsilon_q(p+\Delta)}} \sqrt{\frac{\epsilon_b(p)+
  m_b}{2\epsilon_b(p)}}\cr
&&\times\Biggl\{\frac{({\bf
p\Delta})}{p{\bf
  \Delta}^2}\Biggl(1+\frac{{\bf
    p}^2}{[\epsilon_q(p+\Delta)+m_q][\epsilon_b(p)+m_b]}-
\frac{E_{AV}}{\epsilon_q(p+\Delta)+m_q}\cr
&&
+\frac23\frac{{\bf p}^2}{E_{AV}+
  M_{AV}}\left(\frac1{\epsilon_q(p+\Delta)+m_q}-\frac1{\epsilon_c(p)+m_c}\right)
\Biggl[ \frac{{\bf
    \Delta}^2}{[\epsilon_q(p+\Delta)+m_q][\epsilon_b(p)+m_b]}\cr
&&
+E_{AV}\left(\frac1{\epsilon_q(p+\Delta)+m_q}-\frac1{\epsilon_b(p)+m_b}\right)\Biggr]\Biggr)
+\frac{p}3
\Biggl[\frac{1}{[\epsilon_q(p+\Delta)+m_q][\epsilon_b(p)+m_b]} 
\cr
&&  
-\frac1{E_{AV}}\left(\frac1{\epsilon_q(p+\Delta)+m_q}+\frac1{\epsilon_b(p)+m_b}\right)\Biggr]\Biggr\}\psi_{B_c}({\bf
  p}),
\end{eqnarray}

\begin{eqnarray}
 \label{eq:gv2s}
  g_{V_2}^{S(2)}(q^2)&=&2E_{AV}\sqrt{\frac{E_{AV}}{M_{B_c}}}\int \frac{{\rm
  d}^3p}{(2\pi)^3} \bar\psi_{AV}\left({\bf p}+\frac{2m_c}{E_{AV}+M_{AV}}{\bf
  \Delta} \right)\sqrt{\frac{\epsilon_q(\Delta)+
  m_q}{2\epsilon_q(\Delta)}} \cr
&&\times\Biggl\{\frac{({\bf
p\Delta})}{p{\bf\Delta}^2}\frac{E_{AV}+\epsilon_q(\Delta)-m_q}{\epsilon_q(\Delta)[\epsilon_q(\Delta)+m_q]}
\left[M_{B_c}-\epsilon_b(p)-\epsilon_c(p)\right]\cr
&&
+\frac{p}3\Biggl[\frac1{2\epsilon_q(\Delta)[\epsilon_q(\Delta)+m_q]}\left(\frac1{E_{AV}}+\frac1{\epsilon_q(\Delta)+m_q}\right)+\frac1{4m_b^2}\left(\frac1{E_{AV}}-\frac1{\epsilon_q(\Delta)+m_q}\right)\Biggr]\cr
&&
\Bigl[M_{B_c}+M_{AV}-\epsilon_b(p)-\epsilon_c(p)
-\epsilon_q\left(p+\frac{2m_c}{E_{AV}+M_{AV}}\Delta\right)\cr
&&-\epsilon_c\left(p+\frac{2m_c}{E_{AV}+M_{AV}}\Delta \right)\Bigr] 
-\frac{1}{2m_c\epsilon_q(\Delta)[\epsilon_q(\Delta)+m_q]}\cr
&&
\times\left[M_{AV}-\epsilon_q\left(p+\frac{2m_c}{E_{AV}+M_{AV}}\Delta\right)
-\epsilon_c\left(p+\frac{2m_c}{E_{AV}+M_{AV}}\Delta \right)\right]
\Biggr]\Biggr\}\psi_{B_c}({\bf p}),  
\end{eqnarray}

\begin{eqnarray}
 \label{eq:gv2v}
  g_{V_2}^{V(2)}(q^2)&=&2E_{AV}\sqrt{\frac{E_{AV}}{M_{B_c}}}\int \frac{{\rm
  d}^3p}{(2\pi)^3} \bar\psi_{AV}\left({\bf p}+\frac{2m_c}{E_{AV}+M_{AV}}{\bf
  \Delta} \right)\sqrt{\frac{\epsilon_q(\Delta)+
  m_q}{2\epsilon_q(\Delta)}} \cr
&&\times\frac{p}{6m_c}\Biggl\{\frac{1}{[\epsilon_q(\Delta)+m_q]^2}\left(\frac{E_{AV}}{\epsilon_q(\Delta)}+2\right)\left[M_{B_c}-\epsilon_b(p)-\epsilon_c(p)\right]
-\left(\frac{1}{\epsilon_q(\Delta)+m_q}+\frac1{E_{AV}}\right)\cr
&&\times
\left(\frac{1}{\epsilon_q(\Delta)+m_q}+\frac1{2m_b}\right)
\Bigl[M_{B_c}+M_{AV}-\epsilon_b(p)-\epsilon_c(p)\cr
&&
-\epsilon_q\left(p+\frac{2m_c}{E_{AV}+M_{AV}}\Delta\right)-\epsilon_c\left(p+\frac{2m_c}{E_{AV}+M_{AV}}\Delta
\right)\Bigr]
\Biggr\}\psi_{B_c}({\bf p}),  
\end{eqnarray}

\begin{eqnarray}
  \label{eq:gv3}
  g_{V_3}^{(1)}(q^2)&=&2E_{AV}\sqrt{E_{AV}M_{B_c}}\int \frac{{\rm
  d}^3p}{(2\pi)^3} \bar\psi_{AV}\left({\bf p}+\frac{2m_c}{E_{AV}+M_{AV}}{\bf
  \Delta} \right)\sqrt{\frac{\epsilon_q(p+\Delta)+
  m_q}{2\epsilon_q(p+\Delta)}} \sqrt{\frac{\epsilon_b(p)+
  m_b}{2\epsilon_b(p)}}\cr
&&\times\frac{({\bf
p\Delta})}{p{\bf
\Delta}^2}\Biggl[\frac1{\epsilon_q(p+\Delta)+m_q}-\frac23\frac{{\bf
q}^2}{E_{AV}+ M_{AV}}
\left(\frac1{\epsilon_q(p+\Delta)+m_q}-\frac1{\epsilon_c(p)+m_c}\right)\cr
&&
\times\left(\frac1{\epsilon_q(p+\Delta)+m_q}-\frac1{\epsilon_b(p)+m_b}\right)\Biggr]\psi_{B_c}({\bf p}),  
\end{eqnarray}

\begin{eqnarray}
 \label{eq:gv3s}
  g_{V_3}^{S(2)}(q^2)&=&2E_{AV}\sqrt{E_{AV}M_{B_c}}\int \frac{{\rm
  d}^3p}{(2\pi)^3} \bar\psi_{AV}\left({\bf p}+\frac{2m_c}{E_{AV}+M_{AV}}{\bf
  \Delta} \right)\sqrt{\frac{\epsilon_q(\Delta)+
  m_q}{2\epsilon_q(\Delta)}} \cr
&&\times\left(-\frac{({\bf
p\Delta})}{p{\bf\Delta}^2}\right)\frac{1}{\epsilon_q(\Delta)[\epsilon_q(\Delta)+m_q]}
\left[M_{B_c}-\epsilon_b(p)-\epsilon_c(p)\right]\psi_{B_c}({\bf p}),  
\end{eqnarray}

\begin{eqnarray}
 \label{eq:gv3v}
  g_{V_3}^{V(2)}(q^2)&=&2E_{AV}\sqrt{E_{AV}M_{B_c}}\int \frac{{\rm
  d}^3p}{(2\pi)^3} \bar\psi_{AV}\left({\bf p}+\frac{2m_c}{E_{AV}+M_{AV}}{\bf
  \Delta} \right)\sqrt{\frac{\epsilon_q(\Delta)+
  m_q}{2\epsilon_q(\Delta)}} \cr
&&\times\left(-\frac{p}{6m_c\epsilon_q(\Delta)[\epsilon_q(\Delta)+m_q]^2}\right)\Bigl[M_{B_c}+M_{AV}-\epsilon_b(p)-\epsilon_c(p)\cr
&&
-\epsilon_q\left(p+\frac{2m_c}{E_{AV}+M_{AV}}\Delta\right)-\epsilon_c\left(p+\frac{2m_c}{E_{AV}+M_{AV}}\Delta \right)\Bigr]\psi_{B_c}({\bf p}),  
\end{eqnarray}

\begin{eqnarray}
  \label{eq:ga}
  g_{A}^{(1)}(q^2)&=&(M_{B_c}+M_{AV})\sqrt{\frac{E_{AV}}{M_{B_c}}}\int \frac{{\rm
  d}^3p}{(2\pi)^3} \bar\psi_{AV}\left({\bf p}+\frac{2m_c}{E_{AV}+M_{AV}}{\bf
  \Delta} \right)\sqrt{\frac{\epsilon_q(p+\Delta)+
  m_q}{2\epsilon_q(p+\Delta)}}\cr
&&\times \sqrt{\frac{\epsilon_b(p)+
  m_b}{2\epsilon_b(p)}}
\frac{p}3\Biggl[\frac{1}{E_{AV}+ M_{AV}}
\left(\frac1{\epsilon_q(p+\Delta)+m_q}-\frac1{\epsilon_c(p)+m_c}\right)\cr
&&\times
\Biggl(1-\frac{{\bf
    q}^2}{[\epsilon_q(p+\Delta)+m_q][\epsilon_b(p)+m_b]}\Biggr)+
\frac{1}{[\epsilon_q(p+\Delta)+m_q][\epsilon_b(p)+m_b]}\Biggr]\Biggr\}\psi_{B_c}({\bf
  p}),  \cr
&&
\end{eqnarray}

\begin{eqnarray}
 \label{eq:gas}
  g_{A}^{S(2)}(q^2)&=&(M_{B_c}+M_{AV})\sqrt{\frac{E_{AV}}{M_{B_c}}}\int \frac{{\rm
  d}^3p}{(2\pi)^3} \bar\psi_{AV}\left({\bf p}+\frac{2m_c}{E_{AV}+M_{AV}}{\bf
  \Delta} \right)\sqrt{\frac{\epsilon_q(\Delta)+
  m_q}{2\epsilon_q(\Delta)}} \cr
&&\times\frac{p}{3[\epsilon_q(\Delta)+m_q]}\Biggl\{-\left(\frac{1}{\epsilon_q(\Delta)[\epsilon_q(\Delta)+m_q]}+\frac1{2m_b^2}\right)\Bigl[M_{B_c}+M_{AV}-\epsilon_b(p)-\epsilon_c(p)\cr
&&
-\epsilon_q\left(p+\frac{2m_c}{E_{AV}+M_{AV}}\Delta\right)-\epsilon_c\left(p+\frac{2m_c}{E_{AV}+M_{AV}}\Delta \right)\Bigr]
-\frac{1}{m_b\epsilon_q(\Delta)}
 \cr
&&\times\left[M_{AV}-\epsilon_q\left(p+\frac{2m_c}{E_{AV}+M_{AV}}\Delta\right)
-\epsilon_c\left(p+\frac{2m_c}{E_{AV}+M_{AV}}\Delta \right)\right]
\Biggr\}\psi_{B_c}({\bf p}),  \cr &&
\end{eqnarray}

\begin{eqnarray}
 \label{eq:gav}
  g_{A}^{V(2)}(q^2)&=&(M_{B_c}+M_{AV})\sqrt{\frac{E_{AV}}{M_{B_c}}}\int \frac{{\rm
  d}^3p}{(2\pi)^3} \bar\psi_{AV}\left({\bf p}+\frac{2m_c}{E_{AV}+M_{AV}}{\bf
  \Delta} \right)\sqrt{\frac{\epsilon_q(\Delta)+
  m_q}{2\epsilon_q(\Delta)}} \cr
&&\times\frac{p}{6m_c[\epsilon_q(\Delta)+m_q]}
\left(-\frac1{\epsilon_q(\Delta)+m_q}+\frac1{2m_b}\right)\Bigl[M_{B_c}+M_{AV}-\epsilon_b(p)-\epsilon_c(p)\cr
&&
-\epsilon_q\left(p+\frac{2m_c}{E_{AV}+M_{AV}}\Delta\right)-\epsilon_c\left(p+\frac{2m_c}{E_{AV}+M_{AV}}\Delta \right)\Bigr]\psi_{B_c}({\bf p}),  
\end{eqnarray}

(d) $B_c\to T(^3P_2)$ transition ($T=\chi_{c2},D_2^*$)

\begin{eqnarray}
  \label{eq:tv}
  t_{V}^{(1)}(q^2)&=&(M_{B_c}+M_{T})E_{T}\sqrt{\frac{E_{T}}{M_{B_c}}}\int \frac{{\rm
  d}^3p}{(2\pi)^3} \bar\psi_{T}\left({\bf p}+\frac{2m_c}{E_{T}+M_{T}}{\bf
  \Delta} \right)\sqrt{\frac{\epsilon_q(p+\Delta)+
  m_q}{2\epsilon_q(p+\Delta)}}\cr
&&\times \sqrt{\frac{\epsilon_b(p)+
  m_b}{2\epsilon_b(p)}}\Biggl\{\frac{({\bf
p\Delta})}{p{\bf\Delta}^2}\Biggl[\frac1{\epsilon_q(p+\Delta)+m_q}-\frac13\frac{{\bf
p}^2}{E_{T}+ M_{T}}\cr
&&
\times\left(\frac1{\epsilon_q(p+\Delta)+m_q}-\frac1{\epsilon_c(p)+m_c}\right)\left(\frac1{\epsilon_q(p+\Delta)+m_q}-\frac1{\epsilon_b(p)+m_b}\right)\Biggr]\cr
&&
-\frac{p}{3(E_{T}+ M_{T})[\epsilon_q(p+\Delta)+m_q]}
\left(\frac1{\epsilon_q(p+\Delta)+m_q}-\frac1{\epsilon_c(p)+m_c}\right)\Biggr\}\psi_{B_c}({\bf p}),\cr
&&  
\end{eqnarray}

\begin{eqnarray}
 \label{eq:tvs}
  t_{V}^{S(2)}(q^2)&=&(M_{B_c}+M_{T})E_{T}\sqrt{\frac{E_{T}}{M_{B_c}}}\int \frac{{\rm
  d}^3p}{(2\pi)^3} \bar\psi_{T}\left({\bf p}+\frac{2m_c}{E_{T}+M_{T}}{\bf
  \Delta} \right)\sqrt{\frac{\epsilon_q(\Delta)+
  m_q}{2\epsilon_q(\Delta)}} \cr
&&\times\left(-\frac{({\bf
p\Delta})}{p{\bf\Delta}^2}\right)\frac{1}{\epsilon_q(\Delta)[\epsilon_q(\Delta)+m_q]}
\left[M_{B_c}-\epsilon_b(p)-\epsilon_c(p)\right]\psi_{B_c}({\bf p}),  
\end{eqnarray}

\begin{eqnarray}
 \label{eq:tvv}
  t_{V}^{V(2)}(q^2)&=&0,  
\end{eqnarray}

\begin{eqnarray}
  \label{eq:ta1}
  t_{A_1}^{(1)}(q^2)&=&2\sqrt{E_{T}M_{B_c}}\frac{E_{T}}{M_{B_c}+M_{T}}\int \frac{{\rm
  d}^3p}{(2\pi)^3} \bar\psi_{T}\left({\bf p}+\frac{2m_c}{E_{T}+M_{T}}{\bf
  \Delta} \right)\sqrt{\frac{\epsilon_q(p+\Delta)+
  m_q}{2\epsilon_q(p+\Delta)}}\cr
&&\times \sqrt{\frac{\epsilon_b(p)+
  m_b}{2\epsilon_b(p)}}\Biggl\{\frac{({\bf
p\Delta})}{p{\bf\Delta}^2}\Biggl[1-\frac{{\bf p}^2}{[\epsilon_q(p+\Delta)+m_q][\epsilon_b(p)+m_b]}\Biggr]\cr
&&-\frac13\frac{{\bf
p}^2}{E_{T}+ M_{T}}
\left(\frac1{\epsilon_q(p+\Delta)+m_q}-\frac1{\epsilon_c(p)+m_c}\right)
\cr
&&
\times\Biggl[1+\frac{{\bf p}^2}{[\epsilon_q(p+\Delta)+m_q][\epsilon_b(p)+m_b]}\Biggr]\Biggr\}\psi_{B_c}({\bf p}),\cr
&&  
\end{eqnarray}

\begin{eqnarray}
 \label{eq:ta1s}
  t_{A_1}^{S(2)}(q^2)&=&2\sqrt{E_{T}M_{B_c}}\frac{E_{T}}{M_{B_c}+M_{T}}\int \frac{{\rm
  d}^3p}{(2\pi)^3} \bar\psi_{T}\left({\bf p}+\frac{2m_c}{E_{T}+M_{T}}{\bf
  \Delta} \right)\sqrt{\frac{\epsilon_q(\Delta)+
  m_q}{2\epsilon_q(\Delta)}} \cr
&&\times\Biggl\{\frac{({\bf
p\Delta})}{p{\bf \Delta}^2}\frac{\epsilon_q(\Delta)-m_q}{\epsilon_q(\Delta)[\epsilon_q(\Delta)+m_q]}
\left[M_{B_c}-\epsilon_b(p)-\epsilon_c(p)\right]\cr
&&
+\frac{p}{6[\epsilon_q(\Delta)+m_q]}\Biggl[\Biggl(\frac1{\epsilon_q(\Delta)[\epsilon_q(\Delta)+m_q]}+\frac1{2m_b^2}\Biggr)\Bigl[M_{B_c}+M_{T}-\epsilon_b(p)-\epsilon_c(p)\cr
&&
-\epsilon_q\left(p+\frac{2m_c}{E_{T}+M_{T}}\Delta\right)-\epsilon_c\left(p+\frac{2m_c}{E_{T}+M_{T}}\Delta \right)\Bigr]
+\frac{1}{m_b\epsilon_q(\Delta)}
 \cr
&&\times\left[M_{T}-\epsilon_q\left(p+\frac{2m_c}{E_{T}+M_{T}}\Delta\right)
-\epsilon_c\left(p+\frac{2m_c}{E_{T}+M_{T}}\Delta \right)\right]
\Biggr]\Biggr\}\psi_{B_c}({\bf p}),  
\end{eqnarray}

\begin{eqnarray}
 \label{eq:ta1v}
  t_{A_1}^{V(2)}(q^2)&=&2\sqrt{E_{T}M_{B_c}}\frac{E_{T}}{M_{B_c}+M_{T}}\int \frac{{\rm
  d}^3p}{(2\pi)^3} \bar\psi_{T}\left({\bf p}+\frac{2m_c}{E_{T}+M_{T}}{\bf
  \Delta} \right)\sqrt{\frac{\epsilon_q(\Delta)+
  m_q}{2\epsilon_q(\Delta)}} \cr
&&\times
\frac{p}{3m_c[\epsilon_q(\Delta)+m_q]^2}\Bigl[M_{T}-\epsilon_q\left(p+\frac{2m_c}{E_{T}+M_{T}}\Delta\right)\cr
&&
-\epsilon_c\left(p+\frac{2m_c}{E_{T}+M_{T}}\Delta \right)\Bigr]
\Biggr]\Biggr\}\psi_{B_c}({\bf p}),  
\end{eqnarray}

\begin{eqnarray}
  \label{eq:ta2}
  t_{A_2}^{(1)}(q^2)&=&2E_{T}^2\sqrt{\frac{E_{T}}{M_{B_c}}}\int \frac{{\rm
  d}^3p}{(2\pi)^3} \bar\psi_{T}\left({\bf p}+\frac{2m_c}{E_{T}+M_{T}}{\bf
  \Delta} \right)\sqrt{\frac{\epsilon_q(p+\Delta)+
  m_q}{2\epsilon_q(p+\Delta)}} \sqrt{\frac{\epsilon_b(p)+
  m_b}{2\epsilon_b(p)}}\cr
&&\times\Biggl\{\frac{({\bf
p\Delta})}{p{\bf
  \Delta}^2}\Biggl[\frac1{\epsilon_q(p+\Delta)+m_q}-\frac1{E_{T}}\left(1-\frac{{\bf
    p}^2}{[\epsilon_q(p+\Delta)+m_q][\epsilon_b(p)+m_b]}\right)\cr
&&
-\frac{{\bf p}^2}{E_{T}+
  M_{T}}\left(\frac1{\epsilon_q(p+\Delta)+m_q}-\frac1{\epsilon_c(p)+m_c}\right)
\Biggl( \frac1{\epsilon_q(p+\Delta)+m_q}+\frac1{\epsilon_b(p)+m_b}\cr
&&
-\frac{E_{T}}{[\epsilon_q(p+\Delta)+m_q][\epsilon_b(p)+m_b]}\Biggr)\Biggr]
-\frac{p}{3(E_{T}+ M_{T})}
\left(\frac1{\epsilon_q(p+\Delta)+m_q}-\frac1{\epsilon_c(p)+m_c}\right) 
\cr
&&  
\times\Biggl[\frac1{\epsilon_q(p+\Delta)+m_q}-\frac1{E_{T}}\left(1+\frac{{\bf
    p}^2}{[\epsilon_q(p+\Delta)+m_q][\epsilon_b(p)+m_b]}\right)\Biggr]\Biggr\}\psi_{B_c}({\bf
  p}),
\end{eqnarray}

\begin{eqnarray}
 \label{eq:ta2s}
  t_{A_2}^{S(2)}(q^2)&=&2E_{T}^2\sqrt{\frac{E_{T}}{M_{B_c}}}\int \frac{{\rm
  d}^3p}{(2\pi)^3} \bar\psi_{T}\left({\bf p}+\frac{2m_c}{E_{T}+M_{T}}{\bf
  \Delta} \right)\sqrt{\frac{\epsilon_q(\Delta)+
  m_q}{2\epsilon_q(\Delta)}} \cr
&&\times\Biggl\{-\frac{({\bf
p\Delta})}{p{\bf\Delta}^2}\frac{1}{\epsilon_q(\Delta)[\epsilon_q(\Delta)+m_q]}
\left(1-\frac{\epsilon_q(\Delta)+m_q}{E_{T}}\right)
\left[M_{B_c}-\epsilon_b(p)-\epsilon_c(p)\right]\cr
&&
-\frac{p}{3E_{T}[\epsilon_q(\Delta)+m_q]}\Biggl[\left(\frac1{2\epsilon_q(\Delta)[\epsilon_q(\Delta)+m_q]}+\frac1{4m_b^2}\right)\cr
&&
\times\Bigl[M_{B_c}+M_{T}-\epsilon_b(p)-\epsilon_c(p)
-\epsilon_q\left(p+\frac{2m_c}{E_{T}+M_{T}}\Delta\right)\cr
&&-\epsilon_c\left(p+\frac{2m_c}{E_{T}+M_{T}}\Delta \right)\Bigr] 
+\frac{1}{2m_b\epsilon_q(\Delta)}
\Bigl[M_{T}-\epsilon_q\left(p+\frac{2m_c}{E_{T}+M_{T}}\Delta\right)\cr
&&
-\epsilon_c\left(p+\frac{2m_c}{E_{T}+M_{T}}\Delta \right)\Bigr]
\Biggr]\Biggr\}\psi_{B_c}({\bf p}),  
\end{eqnarray}

\begin{eqnarray}
 \label{eq:ta2v}
  t_{A_2}^{V(2)}(q^2)&=&2E_{T}^2\sqrt{\frac{E_{T}}{M_{B_c}}}\int \frac{{\rm
  d}^3p}{(2\pi)^3} \bar\psi_{T}\left({\bf p}+\frac{2m_c}{E_{T}+M_{T}}{\bf
  \Delta} \right)\sqrt{\frac{\epsilon_q(\Delta)+
  m_q}{2\epsilon_q(\Delta)}} \cr
&&\times\frac{p}{3m_c[\epsilon_q(\Delta)+m_q]^2}\Biggl\{
\frac1{2\epsilon_q(\Delta)}\Bigl[M_{B_c}+M_{T}-\epsilon_b(p)-\epsilon_c(p)
-\epsilon_q\left(p+\frac{2m_c}{E_{T}+M_{T}}\Delta\right)\cr
&&-\epsilon_c\left(p+\frac{2m_c}{E_{T}+M_{T}}\Delta \right)\Bigr] 
-\frac{1}{E_T}
\Bigl[M_{T}-\epsilon_q\left(p+\frac{2m_c}{E_{T}+M_{T}}\Delta\right)\cr
&&
-\epsilon_c\left(p+\frac{2m_c}{E_{T}+M_{T}}\Delta \right)\Bigr]
\Biggr]\Biggr\}\psi_{B_c}({\bf p}),  
\end{eqnarray}

\begin{eqnarray}
  \label{eq:ta3}
  t_{A_3}^{(1)}(q^2)&=&2\sqrt{E_{T}M_{B_c}}E_{T}^2\int \frac{{\rm
  d}^3p}{(2\pi)^3} \bar\psi_{T}\left({\bf p}+\frac{2m_c}{E_{T}+M_{T}}{\bf
  \Delta} \right)\sqrt{\frac{\epsilon_q(p+\Delta)+
  m_q}{2\epsilon_q(p+\Delta)}} \sqrt{\frac{\epsilon_b(p)+
  m_b}{2\epsilon_b(p)}}\cr
&&\times\Biggl\{-\frac{({\bf
p\Delta})}{p{\bf
  \Delta}^2}\frac{{\bf p}^2}{E_{T}+
  M_{T}}\left(\frac1{\epsilon_q(p+\Delta)+m_q}-\frac1{\epsilon_c(p)+m_c}\right)\cr
&&
\times
\frac1{[\epsilon_q(p+\Delta)+m_q][\epsilon_b(p)+m_b]}\Biggr\}\psi_{B_c}({\bf
  p}),
\end{eqnarray}
\begin{equation}
 \label{eq:ta3s}
  t_{A_3}^{S(2)}(q^2)=0,  
\end{equation}
\begin{equation}
 \label{eq:ta3v}
  t_{A_3}^{V(2)}(q^2)=0,  
\end{equation}
where the subscript $q=c,u$ refers to the final active quark, the superscripts ``(1)" and ``(2)" correspond to Figs.~\ref{d1} and
\ref{d2}, $S$ and
$V$ correspond to the scalar and vector potentials of the $q\bar q$-interaction,
and
\[\Delta\equiv \left|{\bf \Delta}\right|=\sqrt{\frac{(M_{B_c}^2+M_F^2-q^2)^2}
{4M_{B_c}^2}-M_F^2},\quad (F=S,AV,AV',T)\]
\[ E_F=\sqrt{M_F^2+{\bf \Delta}^2}, \quad
 \epsilon_Q(p+a
\Delta)=\sqrt{m_Q^2+({\bf p}+a{\bf \Delta})^2} \quad
(Q=b,c,s,u,d). \]

\end{document}